\documentclass[hyper,letterpaper,notoc]{JHEP3}

\usepackage{epsfig}
\usepackage{amsbsy}
\usepackage{varioref}
\usepackage{pifont}
\usepackage{amsmath}

\title{
Ideal Fermion Delocalization\\
 in Five Dimensional Gauge Theories
}

\author{R. Sekhar Chivukula and Elizabeth H. Simmons\\
Department of Physics and Astronomy, Michigan State University\\
East Lansing, MI 48824, USA\\
	E-mail: \email{sekhar@msu.edu, esimmons@msu.edu}}

\author{Hong-Jian He\\
Department of Physics, University of Texas\\
Austin, TX 78712, USA\\
	E-mail: \email{hjhe@physics.utexas.edu}}

\author{Masafumi Kurachi\footnote{Current Address: C.N. Yang Institute for Theoretical
Physics, State University of New York, Stony Brook, NY 11794, USA}\,   and Masaharu Tanabashi\\
Department of Physics, Tohoku University\\
Sendai 980-8578, Japan\\
	E-mail: \email{kurachi@insti.physics.sunysb.edu, tanabash@tuhep.phys.tohoku.ac.jp}}

\abstract{
We discuss ideal delocalization of fermions in a bulk 
$SU(2)\times SU(2) \times U(1)$ Higgsless model with a flat or warped extra dimension.
So as to make an extra dimensional interpretation possible, both the weak and
hypercharge properties of the fermions are delocalized, with
the $U(1)_Y$ current of left-handed fermions being correlated with the $SU(2)_W$
current.   We find that (to subleading order)
ideal fermion delocalization yields vanishing precision electroweak
corrections in this continuum model, as found in corresponding
theory space models based on deconstruction.  In addition to explicit calculations,
we present an intuitive argument for our results based on Georgi's spring analogy.
We also discuss the conditions under which the essential features of an $SU(2)\times
SU(2)\times U(1)$ bulk gauge theory can be captured by a simpler $SU(2) \times
SU(2)$ model.
}


\preprint{MSUHEP-050912\\
TU-752}

\begin{document}


\section{Introduction}

Higgsless models \cite{Csaki:2003dt} have gained popularity because of their ability to provide
an alternative mechanism of electroweak symmetry breaking that forgoes a scalar Higgs  boson 
\cite{Higgs:1964ia}.   Much has been written about
 models \cite{Agashe:2003zs,Csaki:2003zu} based on a five-dimensional
$SU(2) \times SU(2) \times U(1)$ gauge theory in a slice of Anti-deSitter space, in which 
electroweak symmetry breaking is encoded in the boundary conditions of the gauge fields.   The spectrum
includes states identified with the photon, $W$, and $Z$, and also  an infinite tower of  additional massive vector bosons (the higher Kaluza-Klein  or $KK$ excitations), whose exchange is responsible for unitarizing longitudinal $W$ and $Z$ boson scattering \cite{SekharChivukula:2001hz,Chivukula:2002ej,Chivukula:2003kq,He:2004zr}. 

The properties of Higgsless models may be studied 
\cite{Foadi:2003xa,Hirn:2004ze,Casalbuoni:2004id,Chivukula:2004pk,Perelstein:2004sc,Chivukula:2004af,Georgi:2004iy,SekharChivukula:2004mu}
using deconstruction 
\cite{Arkani-Hamed:2001ca,Hill:2000mu} which leads one to 
compute the electroweak parameters $\alpha S$ and $\alpha T$  
\cite{Peskin:1992sw,Altarelli:1990zd,Altarelli:1991fk} in a 
related linear moose model \cite{Georgi:1985hf}. We have shown 
\cite{SekharChivukula:2004mu} how to compute 
all four of the leading zero-momentum electroweak parameters defined
by Barbieri et. al. \cite{Barbieri:2004qk} in a very general class of linear moose models.
We have demonstrated
that a Higgsless model with localized fermions cannot simultaneously satisfy  (1) unitarity bounds, (2) provide acceptably small precision electroweak corrections, and (3) have no light vector bosons other than the photon, $W$, and $Z$.  We also found that localizing the hypercharge properties of the fermions at a single $U(1)$ site adjacent to the chain of $SU(2)$ groups on the linear moose caused $\Delta\rho$ ($Y$ in the language of Barbieri et al. \cite{Barbieri:2004qk}) to vanish.  

Following proposals \cite{Cacciapaglia:2004rb,Foadi:2004ps,Foadi:2005hz} that delocalizing fermions within the extra dimension\footnote{In deconstructed language, delocalization means allowing fermions to derive electroweak properties from more than one site on the lattice of gauge groups \protect\cite{Chivukula:2005bn,Casalbuoni:2005rs}} can reduce electroweak corrections, we showed \cite{SekharChivukula:2005xm}
in an arbitrary Higgsless model that choosing the probability distribution of the delocalized fermions to be related to the wavefunction of the $W$ boson makes the other three ($\hat S$, $\hat T$, $W$) leading zero-momentum precision electroweak parameters  defined by Barbieri, et. al. \cite{Barbieri:2004qk} vanish at tree-level.  We denote such fermions as ``ideally delocalized".   

In this paper, we provide a continuum realization of ideal delocalization that preserves the characteristic of vanishing precision electroweak corrections up to subleading order.  The challenge is as follows.  We have found that deconstructed models with $\Delta\rho = 0$ have the hypercharge current of fermions localized at one site while models with small $\hat S$, $\hat T$ and $W$ have the weak current of fermions ideally delocalized over many sites. This situation is perfectly consistent in  the context of a theory-space moose model, but is difficult to interpret as a model with an extra dimension.  After all, left-handed quarks and leptons carry both $SU(2)$ and $U(1)$ charges, yet should have a single profile along the extra dimension.  

We show here that arranging for the delocalization of the left-handed $U(1)$ fermion current to be correlated with the ideal delocalization of the fermions' $SU(2)$ properties provides a resolution in the context of a bulk  
$SU(2)\times SU(2) \times U(1)$ model.   The moose diagram corresponding to this continuum model is shown in Fig.~\ref{fig:moose_su2sqru1}.  Two species of bulk fermions are introduced.  Fermion $A$ feels the
$SU(2)$ gauge field of the $A$-branch weak groups, while fermion $B$ couples to the $B$-branch fields; 
both couple with the same bulk $U(1)$ gauge field.  By calculating the profiles of the gauge bosons
and fermions, and their couplings to one another, we will demonstrate that ideal fermion delocalization, as realized here in the continuum, still ensures the vanishing of the leading zero-momentum electroweak precision observables -- including $\Delta\rho$.  Moreover, we will find that the essential features of the theory-space model can be captured by an even simpler $SU(2)\times SU(2)$ 
continuum model, which can then be used to study other aspects of the phenomenology of Higgsless models \cite{Chivukula:2005ji}.

Section 2 uses Georgi's spring analogy \cite{Georgi:2004iy} to provide an intuitive understanding of the correspondence  between the $SU(2)^2 \times U(1)$ model and the  $SU(2)^2$ model.  Sections 3 and 4 provide detailed analyses of ideal delocalization in bulk $SU(2)\times SU(2) \times U(1)$ models in flat and warped space, respectively.  The calculation of electroweak observables and is discussed
in section 5. In section 6 we consider the effect of a TeV brane $U(1)$ kinetic energy
term and show that, unlike the case of a Planck brane term, there is no correspondence to an
$SU(2)^2$ model and there are nontrivial electroweak corrections. Section 7 presents our
conclusions.


\section{Mass-Spring Analogy}

\begin{figure}[bt]
  \centering
  \includegraphics[width=8cm]{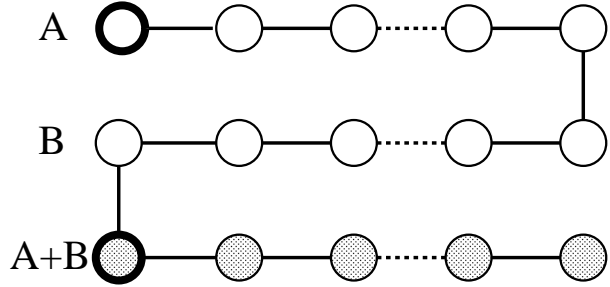}
  \caption{The moose description of the extra dimension model
  discussed in this paper.
  The unshaded and shaded circles represent $SU(2)$ and $U(1)$ gauge
  fields; brane kinetic terms are located at the thick circles. 
  Two species of bulk fermions are introduced: fermion $A$ feels the
  gauge fields of the $A$-branch of $SU(2)$, while fermion $B$ couples to the 
  $B$-branch fields. 
  Both fermions couple to the same bulk $U(1)$ gauge field.
  The delocalization of left-handed $U(1)_Y$ current is therefore correlated with
  the $SU(2)_W$ delocalization. 
  }
  \label{fig:moose_su2sqru1}
\end{figure}

In this paper, we study ideal fermion delocalization in the context of a five-dimensional $SU(2)_A \otimes
SU(2)_B \otimes U(1)$ gauge theory, considering both the case in which the fifth dimension is flat and the case in which it is warped.  The coupling of both $SU(2)$ groups is denoted $g_{5W}$ while that of the hypercharge group is written $g_{5Y}$. We also introduce brane kinetic terms  of strength $g_0$ and $g_Y$ for $SU(2)_A$ and $U(1)$, respectively.  The corresponding moose model is shown in Fig.   \ref{fig:moose_su2sqru1}. 

Explicit analyses presented in sections 3-5 will demonstrate that  all four leading precision electroweak parameters ($\alpha S$, $\alpha T$, $\Delta \rho$, and $\alpha \delta$ \cite{Chivukula:2004af}) vanish to order
$M^2_W/M^2_{W1}$.  It will also be  shown that the $\gamma$, $W$,
and $Z$ couplings and wavefunctions in this model are equivalent  to those in
an effective $SU(2)\times SU(2)$ model with a $U(1)$ brane kinetic energy term of 
appropriate strength.  Before diving into the detailed calculation, we would like to provide an intuitive basis for understanding our results,  using Georgi's spring
analogy \cite{Georgi:2004iy}.  

\begin{figure}[tbhp]
  \centering
  \includegraphics[width=14cm]{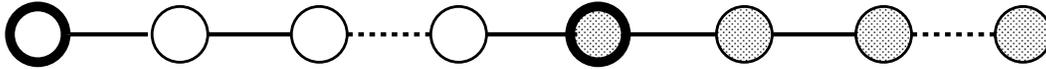}
  \caption{
    Linear moose diagram equivalent to Fig.\protect\ref{fig:moose_su2sqru1}.
    The distribution of a left-handed fermion's $U(1)_Y$ current is correlated with
    the $SU(2)_W$ current.
  }
  \label{fig:moose_su2u1sqr2}
\end{figure}

\begin{figure}[tbhp]
\centering
\includegraphics[width=0.8\textwidth]{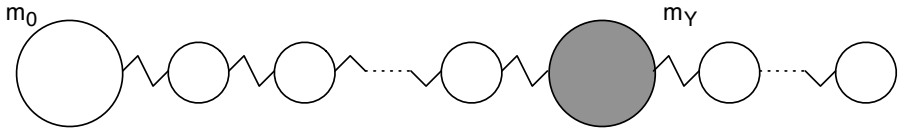}
\label{fig:three}
\caption{Spring system corresponding to the neutral gauge-boson
sector of $SU(2)^2 \times U(1)$ models. $m_0=
1/g^2_0$ and $m_Y = 1/g^2_Y$ are larger than all of the other masses in the chain.}
\end{figure}

The moose for the $SU(2)^2 \times U(1)$ model may be redrawn as shown
in Fig.  \ref{fig:moose_su2u1sqr2}. It is now straightforward to use Georgi's spring analogy
to see that the spring system whose eigenmodes correspond to the neutral
gauge-boson mass eigenstates has the form shown in Fig. 
3. Here we associate each gauge group with a mass $m_i$, each link with a massless
spring with Hooke's law constant $k_i$, and the spring displacements ($x_i$)
with the amplitudes of the corresponding eigenvectors of the gauge-boson mass matrix
using \cite{Georgi:2004iy} the correspondence\footnote{This correspondence is easy to see
by comparing the potential energy of a spring system, $\sum_i k_i (x_i-x_{i-1})^2/2$,
with the quadratic form associated with the gauge boson masses, $\sum_i
f^2_i (g_i A^\mu_i - g_{i-1} A^\mu_{i-1})^2/8$.}
\begin{equation}
x_i  \leftrightarrow g_i A^\mu_i~,\label{eq:gA}\quad\quad\quad
k_i  \leftrightarrow \frac{f^2_i}{4}~, \quad\quad\quad
m_i  \leftrightarrow \frac{1}{g^2_i}~.
\end{equation}
Under this correspondence, the mass-squareds of the gauge-boson eigenstates correspond to
the squared frequencies of the spring system, and the eigenmodes of the spring system correspond to the amplitudes of the corresponding gauge-boson eigenstates \cite{Georgi:2004iy}.

Let us first consider the case of a flat extra dimension. In order to obtain light $W$ and
$Z$ bosons, we work in the limits
\begin{equation}
\frac{1}{g^2_Y}  \gg \frac{\pi R}{g^2_{5Y}}~ \quad\quad
\frac{1}{g^2_0}  \gg  \frac{2\pi R}{g^2_{5W}}
\label{eq:condflat}
\end{equation}
and therefore the corresponding masses $m_0=1/g^2_0$ and $m_Y =
1/g^2_Y$ in Fig.  3 are drawn as large. These conditions are required in order
that the $W$ and $Z$ be much lighter than the $KK$ resonances in these
models \cite{Georgi:2004iy,SekharChivukula:2004mu}.

Consider the neutral gauge boson sector. 
Using our physical intuition, it is easy to see which spring system eigenmodes correspond
to the photon and the $Z$-boson. The photon, which is massless, corresponds to
the uniform translation of the spring system -- {\it i.e.} to a ``flat"
gauge-boson profile (see eqn. (\ref{eq:gA})) in which $g_i A_i$ is constant. Since there
is no restoring force for this motion, this corresponds to a zero mode and therefore a massless
photon.

\begin{figure}[bt]
\centering
\includegraphics[width=0.6\textwidth]{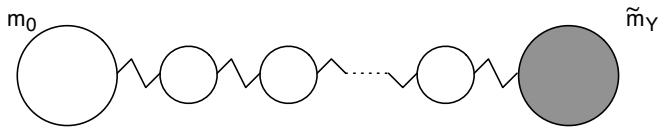}
\label{fig:four}
\caption{Spring system corresponding to the neutral gauge-boson
sector of an $SU(2)^2$ model with $\tilde{m_Y}$ an effective mass
as in eqn. (\protect\ref{eq:tildem}).}
\end{figure}

If $m_0$ and $m_Y$ are much larger than all of the other masses, 
the next lightest (low-frequency) mode corresponds to a ``breathing mode" in which masses
$m_0$ and $m_Y$ oscillate slowly opposite to one another. In this mode, the masses in the
spring system between $m_0$ and $m_Y$ oscillate adiabatically, but the masses in the
chain to the right of $m_Y$ oscillate uniformly with no relative motion. 
That is, the masses in the chain to the right of $m_Y$ have, again, a flat profile. To
leading order, the only effect of the masses to the right of $m_Y$ is to change the
effective mass of $m_Y$ to
\begin{equation}
\tilde{m}_Y = m_Y + \sum_{U(1)\, chain} m_\ell~,
\label{eq:tildem}
\end{equation}
where the sum extends over all of the masses to the right of $m_Y$ -- all
masses in the ``$U(1)$ chain."  Corrections to this picture due to oscillatory motion
within the $U(1)$ chain, will be suppressed by $m^2_\ell/\tilde{m}^2_Y$. 
To this order, therefore, the properties of the oscillatory modes corresponding to both the
photon and $Z$ are equivalent to those calculated in the spring system shown
in Fig.  4.  But this spring system  corresponds to the neutral gauge-boson sector
of an $SU(2)^2$ linear moose in which, as suggested by eqn. (\ref{eq:tildem}),
the strength $g_{Y{\rm eff}}$ of the hypercharge brane kinetic energy term is taken to satisfy the relation
\begin{equation}
\frac{1}{g^2_{Y{\rm eff}}}  =  \frac{1}{g^2_Y} + \frac{\pi R}{g^2_{5Y}}~.
\end{equation}

The properties of the $W$ boson and the charged KK resonances $W_{(n\geq 1)}$ are
independent of the $U(1)$ portion of the moose.  These charged eigenmodes are therefore
identical in the $SU(2)^2\otimes U(1)$ and $SU(2)^2$ linear mooses discussed here.
In addition, we note that the contribution of the $SU(2)^2$ groups to the properties of the neutral
gauge bosons is the same in both models.  

Finally, we may consider the corresponding situation in a Higgsless model
in warped space. At first sight, the situation here appears to be different: the common
deconstruction of this model has a single gauge coupling and geometrically
varying $f$-constants \cite{Falkowski:2002cm,Randall:2002qr}. However,
one may choose an alternative `f-flat' deconstruction 
\cite{SekharChivukula:2005xm,newpaper} in which the couplings vary but the
$f$-constants do not. In this alternate deconstruction, the analysis given above for
a flat space model applies directly.

In sections 3 and 4 of this paper, we will see that the physical intuition just presented is
born out by  explicit calculations of the properties of the photon, $W$, and $Z$ bosons
in a bulk $SU(2)^2 \times U(1)$ Higgsless model of electroweak symmetry breaking.
Moreover, because the precision electroweak corrections $\alpha S$, $\alpha T$, $\Delta\rho$ and $\alpha \delta$ measure the degree to which the $W$ and $Z$ bosons of a given model differ from those of the Standard Model, we expect that the values of these precision observables will be the same, to leading order, for the $SU(2)^2 \otimes U(1) $ and $SU(2)^2$ models discussed here.  Again, we will find, in section 5, that this is supported by explicit calculations.  Note that this agreement occurs despite the fact that the profiles of the individual higher neutral KK resonances will be different in the two models.


\section{Explicit Calculations in Flat Space}

We consider a five-dimensional $SU(2)_A \otimes
SU(2)_B \otimes U(1)$ gauge theory in flat space, in which the fifth dimension
(denoted by the coordinate $y$) is compactified on an interval
of length $\pi R$.
In order to make $M_W$ and $M_Z$ sufficiently lighter than the other
KK masses, we also introduce kinetic terms for $SU(2)_A$ and $U(1)$ on the $y=0$ brane. 
The continuum 5D action corresponding to
Fig.~\ref{fig:moose_su2sqru1} is then given by
\begin{eqnarray}
  S &=& \int_0^{\pi R} dy \int d^4 x \left\{
    -\frac{1}{4g_0^2} \delta(y-0^+) W^a_{A\mu\nu} W^a_{A\rho\lambda}
     \eta^{\mu\rho}\eta^{\nu\lambda}
    -\frac{1}{4g_Y^2} \delta(y-0^+) B_{\mu\nu} B_{\rho\lambda}
     \eta^{\mu\rho}\eta^{\nu\lambda}
    \right\}
    \nonumber\\
    & & 
       +\int_0^{\pi R} dy \int d^4 x \dfrac{1}{g_{5WA}^2} \left\{
    -\frac{1}{4} W^a_{A\mu\nu} W^a_{A\rho\lambda}
     \eta^{\mu\rho}\eta^{\nu\lambda}
    +\frac{1}{2} W^a_{A\mu y} W^a_{A\nu y}
     \eta^{\mu\nu}
    \right\}
    \nonumber\\
    & & 
       +\int_0^{\pi R} dy \int d^4 x \dfrac{1}{g_{5WB}^2} \left\{
    -\frac{1}{4} W^a_{B\mu\nu} W^a_{B\rho\lambda}
     \eta^{\mu\rho}\eta^{\nu\lambda}
    +\frac{1}{2} W^a_{B\mu y} W^a_{B\nu y}
     \eta^{\mu\nu}
    \right\}
    \nonumber\\
    & & 
       +\int_0^{\pi R} dy \int d^4 x \dfrac{1}{g_{5Y}^2} \left\{
    -\frac{1}{4} B_{\mu\nu} B_{\rho\lambda}
     \eta^{\mu\rho}\eta^{\nu\lambda}
    +\frac{1}{2} B_{\mu y} B_{\nu y}
     \eta^{\mu\nu}
    \right\},
\end{eqnarray}
with $W^a_{A\mu}$ ($W^a_{B\mu}$) being
the $SU(2)$ gauge fields in the $A$-($B-$)branches and $B_\mu$ being the
$U(1)$ gauge field.
These gauge fields satisfy boundary conditions,
\begin{equation}
 \partial_y W^a_{A\mu} = 0, \qquad
  W^{1,2}_{B\mu} = 0, \qquad
  W^{3}_{B\mu} = B_{\mu}, \qquad
  \dfrac{1}{g_{5WB}^2} \partial_y  W^{3}_{B\mu} 
  + \dfrac{1}{g_{5Y}^2} \partial_y  B_{\mu} 
  = 0,
\end{equation}
at $y=0$, which break the original gauge group to $SU(2)_W \times U(1)_Y$, and boundary conditions 
\begin{equation}
  W^a_{A\mu} = W^a_{B\mu}, \qquad
  \dfrac{1}{g_{5WA}^2} \partial_y  W^a_{A\mu}
  + \dfrac{1}{g_{5WB}^2} \partial_y  W^a_{B\mu} = 0, 
\qquad
  \partial_y  B_{\mu} = 0, 
\end{equation}
at $y=\pi R$, which break $SU(2)_A \times SU(2)_B$ to its diagonal subgroup.
For simplicity, in the following analyses, 
we assume the bulk $SU(2)$ gauge couplings in the $A$ and $B$
branches are identical:
$  g_{5W}^2 = g_{5WA}^2 = g_{5WB}^2$.

We can unfold the original moose of Fig.~\ref{fig:moose_su2sqru1}, to 
obtain an equivalent linear moose model as shown in Fig.~\ref{fig:moose_su2u1sqr2}.
The corresponding continuum action is 
\begin{eqnarray}
  S &=& \int_0^{3\pi R} \!\! dy \int d^4 x \left\{
    -\frac{1}{4g_0^2} \delta(y-0^+) W^a_{\mu\nu} W^a_{\rho\lambda}
     \eta^{\mu\rho}\eta^{\nu\lambda}
    -\frac{1}{4g_Y^2} \delta(y-2\pi R-0^+) B_{\mu\nu} B_{\rho\lambda}
     \eta^{\mu\rho}\eta^{\nu\lambda}
    \right\}
    \nonumber\\
    & & 
       +\int_0^{2 \pi R} dy \int d^4 x \dfrac{1}{g_{5W}^2} \left\{
    -\frac{1}{4} W^a_{\mu\nu} W^a_{\rho\lambda}
     \eta^{\mu\rho}\eta^{\nu\lambda}
    +\frac{1}{2} W^a_{\mu y} W^a_{\nu y}
     \eta^{\mu\nu}
    \right\}
    \nonumber\\
    & & 
       +\int_{2\pi R}^{3\pi R} dy \int d^4 x \dfrac{1}{g_{5Y}^2} \left\{
    -\frac{1}{4} B_{\mu\nu} B_{\rho\lambda}
     \eta^{\mu\rho}\eta^{\nu\lambda}
    +\frac{1}{2} B_{\mu y} B_{\nu y}
     \eta^{\mu\nu}
    \right\},
\label{eq:5D_straight}
\end{eqnarray}
with boundary conditions 
\begin{eqnarray}
 {\rm at\ } y&=&0:\ \ \ \  \qquad\qquad \partial_y W^a_{\mu} = 0\\
{\rm at\ } y &=& 2\pi R: \qquad\qquad 
  W^{1,2}_\mu = 0, \qquad
  \dfrac{1}{g_{5W}^2} \partial_y W^3_\mu =
  \dfrac{1}{g_{5Y}^2} \partial_y B_\mu\\
 {\rm at\ } y &=& 3\pi R: \qquad\qquad  
  \partial_y B_{\mu} = 0.
\end{eqnarray}

For the Fig.~\ref{fig:moose_su2u1sqr2} linear moose, 
the weak $SU(2)$ current distribution of a fermion,\footnote{In practice a current distribution
for the ordinary fermions means that the observed fermions are the lightest eigenstates of 
five-dimensional fermions, just as the  $W$ and $Z$ gauge-bosons are the lightest in a tower of 
``KK" excitations \protect\cite{Cacciapaglia:2004rb}. The fermion wavefunction is the 
wavefunction for this lightest eigenstate.}  $|\psi(y)|_W^2$, is
defined on $y \in [0,2\pi R)$, while the left-handed $U(1)$ hypercharge
distribution, $|\psi(y)|_Y^2$, takes its value on $y \in [2\pi R,3\pi R)$.
These current distributions are
correlated, as a consequence of the original folded structure 
(Fig.~\ref{fig:moose_su2sqru1}):
\begin{equation}
  |\psi(y+2\pi R)|_Y^2 = |\psi(y)|_W^2 + |\psi(2\pi R - y)|_W^2 , 
\qquad
  \mbox{for $\pi R \ge y \ge 0$}.
\label{eq:u1delocalize}
\end{equation}
This observation is what makes it possible to generalize our 
results \cite{SekharChivukula:2005xm,Chivukula:2005ji} for
theory-space models with ideal fermion delocalization to five-dimensional gauge theories.

\subsection{Mode equations and modified BCs}

The 5D fields $W^a_\mu(x,y)$ and $B_\mu(x,y)$ can be decomposed into
KK-modes,
\begin{eqnarray}
  W_\mu^{1,2}(x,y) &=& \sum_{n} W_\mu^{(n)1,2}(x) \chi_{W(n)}(y),\\  
  W_\mu^3(x,y) &=& \gamma_\mu(x) \chi_\gamma(y)
    +\sum_{n} Z_\mu^{(n)} \chi_{Z(n)}(y), \qquad \mbox{for $y<2\pi R$},
  \\
  B_\mu(x,y) &=& \gamma_\mu(x) \chi_\gamma(y)
    +\sum_{n} Z_\mu^{(n)} \chi_{Z(n)}(y), \qquad \mbox{for $y>2\pi R$}.
\end{eqnarray}
Here $\gamma_\mu(x)$ is the photon, and $W_\mu^{(n)\, 1,2}(x)$ and 
$Z_\mu^{(n)}(x)$ are the $KK$ towers of the massive $W$ and $Z$ bosons,
the lowest of which correspond to the observed $W$ and $Z$ bosons. 
Since the lightest massive $KK$-modes are identified as the observed $W$ and $Z$ bosons,
we write
\begin{eqnarray}
 M_W^2 &\equiv& M_{W_{(0)}}^2,\ \ \ \ \  M_Z^2 \equiv M_{Z_{(0)}}^2~,\\
 \chi_W &\equiv&  \chi_{W_{(0)}}~,\ \ \ \ \  \chi_Z \equiv 
  \chi_{Z_{(0)}}~.
\end{eqnarray}

These mode functions  obey differential equations derived from the 5D
Lagrangian eqn. (\ref{eq:5D_straight}):
\begin{eqnarray}
  & &  
  0 = \partial_y^2 \chi_W(y) + M_W^2 \chi_W(y), 
  \label{eq:mode_eq_W}
  \\
  & &
  0 = \partial_y^2 \chi_Z(y) + M_Z^2 \chi_Z(y), 
  \label{eq:mode_eq_Z}
  \\
  & &
  0 = \partial_y^2 \chi_\gamma(y).
  \label{eq:mode_eq_gamma}
\end{eqnarray}
which  hold for $0^+ < y < 2\pi R$ and 
$2\pi R +0^+ < y < 3\pi R$.
The presence of brane kinetic terms is reflected in modifications of the boundary
conditions. We find
\begin{equation}
  0 = \left.
    \partial_y \chi_W(y)\right|_{y=0^+} + \dfrac{g_{5W}^2}{g_0^2} M_W^2 \left.\chi_W(y)\right|_{y=0}, \ \ \ \ \ 
  0 = \left. \chi_W(y)\right|_{y=2\pi R},
\label{eq:bcW2}
\end{equation}
for the $W$ mode function,
\begin{equation}
  0 = 
    \left. \partial_y \chi_Z(y)\right|_{y=0^+} + \dfrac{g_{5W}^2}{g_0^2} M_Z^2 \left.\chi_Z(y)
  \right|_{y=0},\ \ \ \ \ 
\left.  \chi_Z(y)\right|_{y=2\pi R-0^+}= \left. \chi_Z(y)\right|_{y=2\pi R+0^+} , \label{eq:bcZ2}
\end{equation}
\begin{equation}
   \dfrac{1}{g_{5W}^2} \partial_y \left. \chi_Z(y) \right|_{y=2\pi R-0^+}
\ \   = \ \   \dfrac{1}{g_{5Y}^2} \partial_y \left. \chi_Z(y) \right|_{y=2\pi
    R+0^+} 
   + \dfrac{1}{g_Y^2} M_Z^2 \left. \chi_Z(y)\right|_{y=2\pi R},
\label{eq:bcZ3}
\end{equation}
\begin{equation}
  0 = \left.
    \partial_y \chi_Z(y) 
  \right|_{y=3\pi R}, \label{eq:bcZ4}
\end{equation}
for the $Z$ mode function, and
\begin{equation}
  0= \left. \partial_y \chi_\gamma(y)\right|_{y=0}
\ \ \ \ \ \ 
 \left.  \chi_\gamma(y)\right|_{y=2\pi R-0^+} = \left. \chi_\gamma(y) \right|_{y=2\pi R+0^+},  \ \ \ \ \ \   0 = \left. \partial_y \chi_\gamma(y)\right|_{y=3\pi R}  
\end{equation}
\begin{equation}
  \dfrac{1}{g_{5W}^2} 
  \left. \partial_y \chi_\gamma(y)\right|_{y=2\pi R-0^+}
  = 
  \dfrac{1}{g_{5Y}^2} 
  \left. \partial_y \chi_\gamma(y)\right|_{y=2\pi R+0^+}.
\end{equation}
for the photon.

In order to obtain canonically normalized 4D fields ($W$, $Z$, $\gamma$), 
these mode functions are normalized as
\begin{eqnarray}
  & &
  1 = \int_0^{2\pi R} \!\! dy \left\{
      \dfrac{1}{g_{5W}^2} 
      +\dfrac{\delta(y-0^+)}{g_0^2} 
      \right\}
      \left|\chi_W(y)\right|^2, 
  \label{eq:normW}
  \\
  & &
  1 = \int_0^{3\pi R} \!\! dy \left\{
      \dfrac{\theta(2\pi R-y)}{g_{5W}^2} 
      +\dfrac{\delta(y-0^+)}{g_0^2} 
      +\dfrac{\theta(y-2\pi R)}{g_{5Y}^2} 
      +\dfrac{\delta(y-2\pi R)}{g_Y^2} 
      \right\} \left|\chi_Z(y)\right|^2,
  \nonumber\\
  & &
  \label{eq:normZ}
  \\
  & &
  1 = \int_0^{3\pi R} \!\! dy \left\{
      \dfrac{\theta(2\pi R-y)}{g_{5W}^2} 
      +\dfrac{\delta(y-0^+)}{g_0^2} 
      +\dfrac{\theta(y-2\pi R)}{g_{5Y}^2} 
      +\dfrac{\delta(y-2\pi R)}{g_Y^2} 
      \right\} \left|\chi_\gamma(y)\right|^2.
  \label{eq:normgamma}
  \nonumber\\
  & &
\end{eqnarray}

\subsection{Gauge Boson and Fermion Profiles}

\begin{figure}[htbp]
  \centering
  \includegraphics[width=14cm]{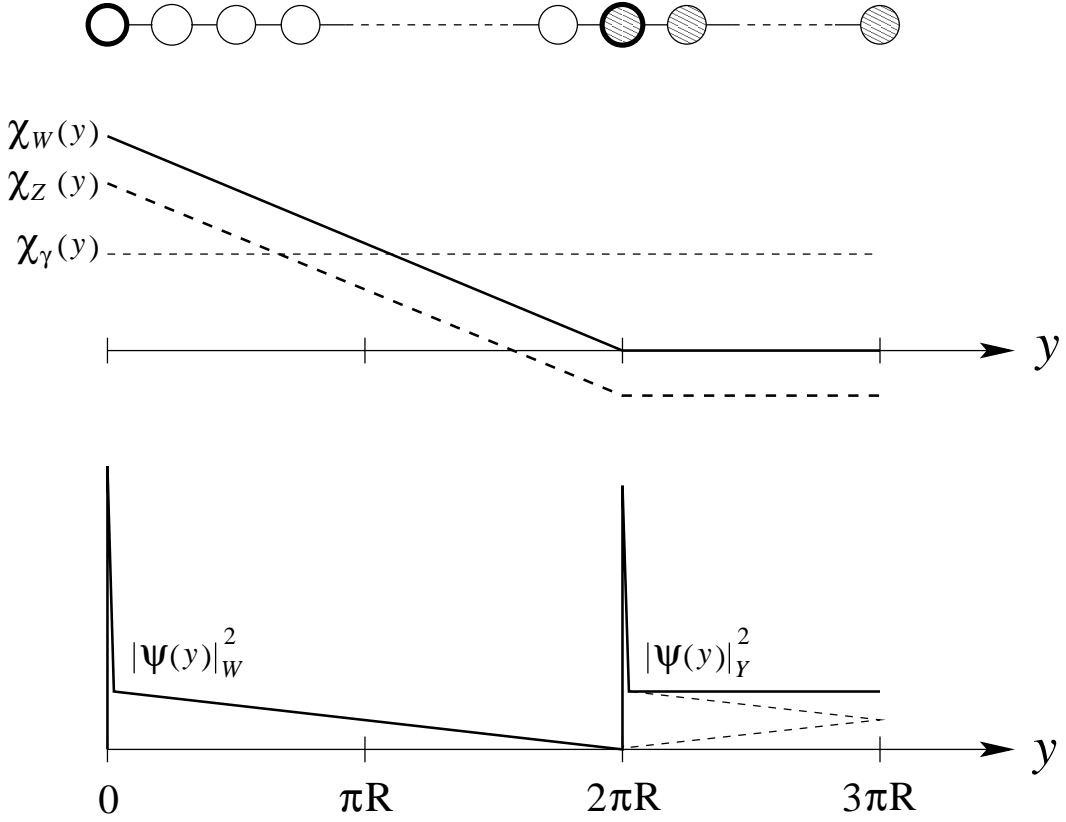}
  \caption{
    Sketch of the gauge boson mode functions and ideally delocalized fermion probability functions for the linear moose model.  Note that, as seen from the text, \protect{$\chi_Z(0) \simeq c\, \chi_W(0)$} and \protect{$\chi_\gamma(0) \simeq s\, \chi_W(0)$} while \protect{$\chi_Z(y)$} is roughly parallel to \protect{$\chi_W(y)$}.  In addition, the fermion profile is flat in the hypercharge region \protect{$2\pi R < y < 3\pi R$}.
  }
  \label{fig:profile}
\end{figure}

We now derive explicit expressions for the profiles of the gauge bosons and ideally delocalized
fermions in the fifth dimension.   To help the reader obtain an intuitive feel for the shapes of the
wavefunctions, they are sketched in Fig. \ref{fig:profile}.

\subsubsection{$W$ profile}

The Dirichlet condition (\ref{eq:bcW2}) at $y=2\pi R$  and the mode
equation (\ref{eq:mode_eq_W}) determine the form of the $W$ mode 
function $\chi_W$,
\begin{equation}
  \chi_W(y) \propto \sin\left[
    M_W (2\pi R - y)
  \right]. 
\label{eq:Wprofile0}
\end{equation}
Because we are interested in situations where $W$ has a much lighter mass 
than the compactification scale $R^{-1}$,
\begin{equation}
  M_W \ll \dfrac{1}{\pi R}.
\end{equation}
we may expand eqn. (\ref{eq:Wprofile0}) in terms of $M_W$,
\begin{equation}
  \chi_W(y) = C_W \left(1-\dfrac{y}{2\pi R}\right)
    \left[1 - \dfrac{4}{3!} \left(M_W \pi R \right)^2 
                            \left(1-\dfrac{y}{2\pi R}\right)^2
    + \cdots
    \right],
    \label{eq:chiwy}
\end{equation} 
with $C_W$ being a normalization constant.
The boundary condition (\ref{eq:bcW2}) at $y=0$  then determines the size of the 
brane kinetic term, $g_0^2$, as a function of $M_W$:
\begin{equation}
  g_0^2 = 2 g_{5W}^2 M_W^2 \pi R
    \left[ 1 + \frac{4}{3} (M_W \pi R)^2 + \cdots \right],
\label{eq:g0vsMw}
\end{equation}
which enables us to find the normalization
constant $C_W$ from eqn. (\ref{eq:normW}):
\begin{equation}
  C_W^2 = 2g_{5W}^2 M_W^2 \pi R 
  \left[
    1 + \dfrac{4}{3} (M_W \pi R)^2 + \cdots
  \right].
\end{equation}

\subsubsection{$Z$ profile}

From the boundary condition (\ref{eq:bcZ2}) at $y=0$, we know the
slope of $\chi_Z(y)$ at $y=0$.  Higher derivative terms of $\chi_Z(y)$
at $y=0$ can also be calculated by using the mode equation
(\ref{eq:mode_eq_Z}). 
Taylor expansion near $y=0$ then gives
\begin{eqnarray}
  \chi_Z(y) = C_Z \left[
    1 - \dfrac{g_{5W}^2}{g_0^2} M_Z^2 y 
      - \dfrac{1}{2}M_Z^2 y^2 
      + \dfrac{1}{3!} \dfrac{g_{5W}^2}{g_0^2} M_Z^4 y^3 +\cdots
  \right],
\label{eq:chi_Z1}
\end{eqnarray}
where $C_Z$ is a normalization constant.
We note that this Taylor expansion can be viewed as an expansion in
terms of $M_Z^2$, and also that $g_0^2$ is given as a function of $M^2_W$ in 
eqn. (\ref{eq:g0vsMw}). 
Similar analysis can be done at $y=3\pi R$, where the slope of
$\chi_Z$ vanishes thanks to the Neumann condition in eqn. (\ref{eq:bcZ4}).
Taylor expansion around $y=3\pi R$ gives
\begin{equation}
  \chi_Z(y) = \hat{C}_Z \left[
    1 - \frac{1}{2} M_Z^2 (3\pi R - y)^2 + \cdots 
  \right].
\end{equation}
The continuity condition (\ref{eq:bcZ2}) at $y=2\pi R$ determines the ratio of
constants $C_Z$ and $\hat{C}_Z$:
\begin{equation}
  \hat{C}_Z 
  = C_Z \left[
        1 - 2 \dfrac{g_{5W}^2}{g_0^2} M_Z^2 \pi R
      - \dfrac{3}{2} M_Z^2 (\pi R)^2
      + \dfrac{1}{3} \dfrac{g_{5W}^2}{g_0^2} M_Z^4 (\pi R)^3
      + \cdots
  \right]~,
  \label{eq:czprime}
\end{equation}
while  eqn. (\ref{eq:bcZ3}) yields the
size of the hypercharge brane kinetic term:
\begin{equation}
  g_Y^2 = 2 g_{5W}^2 (M_Z^2 - M_W^2) \pi R
   \left[ 1 + \frac{4}{3} (M_Z^2 - 2M_W^2) (\pi R)^2
            + 2 \dfrac{g_{5W}^2}{g_{5Y}^2} (M_Z^2 - M_W^2) (\pi R)^2
            + \cdots
   \right].
\label{eq:gYvsM}
\end{equation}
From this result, we may derive an expression for $g_{Y\rm eff}$
\begin{equation}
 \frac{1}{g_{Y{\rm eff}}^2}
  = \frac{1}{g_Y^2} + \frac{\pi R}{g_{5Y}^2}
  = \frac{1}{2g_{5W}^2 (M_Z^2-M_W^2)\pi R}\left(
     1 - \frac{4}{3}(M_Z^2 - 2 M_W^2)(\pi R)^2 + \cdots
    \right). 
\label{eq:gYvsMi}
\end{equation}

We are now ready to determine the normalization constant $C_Z$ from
eqn. (\ref{eq:normZ}).  Initially, it appeared that $C_Z$ might depend on
the bulk $U(1)$ gauge coupling because of the non-trivial dependence
on $g_{5Y}$ in the third term of eqn. (\ref{eq:normZ}).  However, 
the fourth term in eqn. (\ref{eq:normZ}) also depends implicitly on $g_{5Y}$
through $g_Y^2$ (see eqn. (\ref{eq:gYvsM})) and
we find that the $g_{5Y}$ dependence in the two terms cancels at 
the order to which we are working.   By examining the power 
counting in $M_{W,Z}^2$, we see that we can
ignore the $y$ dependence of $\chi_Z$ for $y>2\pi R$  in eqn. (\ref{eq:normZ}) once
eqn. (\ref{eq:gYvsM}) is applied.  Performing the integral yields 
\begin{eqnarray}
\lefteqn{
    \int_0^{3\pi R} \!\! dy \left\{
      \dfrac{\theta(y-2\pi R)}{g_{5Y}^2} 
      +\dfrac{\delta(y-2\pi R)}{g_Y^2} 
    \right\} \left|\chi_Z(y)\right|^2
} \nonumber\\
  &\simeq& \left(\dfrac{\pi R}{g_{5Y}^2} + \dfrac{1}{g_Y^2}\right) 
      \left|\chi_Z(y=2\pi R)\right|^2
  \nonumber\\
  &=& \dfrac{1}{2 g_{5W}^2 (M_Z^2 - M_W^2) \pi R}
      \left(
        1 - \frac{4}{3} (M_Z^2 - 2M_W^2) (\pi R)^2
        + \cdots
      \right) \left|\chi_Z(y=2\pi R)\right|^2
\end{eqnarray}
which confirms the cancellation of all $g_{5Y}^2$ dependence at this order. 
After a straightforward calculation, we obtain the normalization
constant $C_Z$ 
\begin{equation}
  C_Z^2 = 2 g_{5W}^2 \dfrac{M_W^4}{M_Z^2} \pi R \left[
    1 + \dfrac{4}{3}(M_Z^2 - M_W^2)(\pi R)^2 + \cdots \right]~.
\label{eq:cz}
\end{equation}
Note that in eqns. (\ref{eq:czprime}) and (\ref{eq:cz}) neither hypercharge coupling
($g_Y$ or $g_{5Y}$) appears explicitly to this order -- all dependence on these parameters
has been absorbed into $M^2_Z$.

\subsubsection{Photon profile}

The photon mode function possesses a flat profile,
\begin{equation}
  \chi_\gamma(y) = C_\gamma, 
\end{equation}
with $C_\gamma$ being a normalization constant.
The normalization condition eqn. (\ref{eq:normgamma}) then reads
\begin{equation}
  1 = C_\gamma^2 \left(
    \dfrac{2\pi R}{g_{5W}^2} +\dfrac{1}{g_0^2} 
  + \dfrac{\pi R}{g_{5Y}^2}
   + \dfrac{1}{g_Y^2}
    \right).
\end{equation}
Inserting eqn. (\ref{eq:g0vsMw}) and eqn. (\ref{eq:gYvsMi}) into this
expression, we again observe the cancellation of the $g^2_Y$ and $g_{5Y}^2$ dependence
between the third and the fourth terms.
We find
\begin{equation}
  C_\gamma^2 = 2 g_{5W}^2 M_W^2 \pi R 
    \left(1-\dfrac{M_W^2}{M_Z^2}\right)
    \left(
      1-\dfrac{4}{3}(M_W \pi R)^2
     +\cdots
    \right).
\end{equation}

\subsubsection{Ideally delocalized fermions}

We now introduce the wavefunction assumed for the left-handed components of the
ordinary fermions in this 
model.  We focus here on 
ideally-delocalized fermions as defined in \cite{SekharChivukula:2005xm}, which have been found, in
theory-space Higgsless models, to yield small precision electroweak corrections.  
An ideally-delocalized fermion's weak $SU(2)$ current distribution on $y \in [0,2\pi R)$ is derived from the $W$-boson profile:
\begin{equation}
  |\psi(y)|^2_W \propto 
  \left( \dfrac{1}{g_0^2} \delta(y-0^+) + \dfrac{1}{g_{5W}^2} \right)
  \chi_W(y).
  \label{eq:fermprof}
\end{equation}
with the following normalization condition:
\begin{equation}
  1 = \int_0^{2\pi R} dy\, |\psi(y)|^2_W. 
\end{equation}
We thus obtain the ideally delocalized current distribution 
\begin{equation}
  |\psi(y)|^2_W = 
    \left(1-2(M_W \pi R)^2\right) \delta(y - 0^+)
  +M_W^2 (2\pi R - y )  + \cdots,
\end{equation}
where we have neglected terms of order $(M_W \pi R)^4$.

The $U(1)$ current distribution is defined on $2\pi R \le y \le
3\pi R$.  Recalling that it is correlated with the $SU(2)$ fermion profile, as in eqn. (\ref{eq:u1delocalize}), we obtain
\begin{eqnarray}
  |\psi(y)|^2_Y &=& 
  \left(1-2(M_W \pi R)^2\right) \delta(y - 2\pi R - 0^+)
  +M_W^2 (4\pi R - y )  
  +M_W^2 (y - 2\pi R)
  + \cdots
  \nonumber\\
  &=& 
  \left(1-2(M_W \pi R)^2\right) \delta(y - 2\pi R - 0^+)
  +2 M_W^2 \pi R 
  + \cdots
\label{eq:hypercurrentdist}
\end{eqnarray}
which is flat in the bulk.

The right-handed components of the ordinary fermions couple to hypercharge
(and, therefore, to electric charge), and their couplings depend on the wavefunction
for the right-handed components $|\widetilde{\psi}(y)|^2_Y$. This wavefunction
is normalized
\begin{equation}
  1 = \int_{2\pi R}^{3\pi R} dy |\widetilde{\psi}(y)|^2_Y. 
\label{eq:rhnorm}
\end{equation}
We will assume, in what follows, that the left- and right-handed component
wavefunctions satisfy
\begin{equation}
|\psi(y)|^2_Y - |\widetilde{\psi}(y)|^2_Y = {\cal O}\left(M^2_W \pi R\right)~.
\label{eq:lhrhcondition}
\end{equation}
From equation (\ref{eq:hypercurrentdist}) we see that, for example,  a right-handed fermion
wavefunction localized at $y=2\pi R$
\begin{equation}
|\widetilde{\psi}(y)|^2_Y = \delta(y-2\pi R -0^+)~,
\end{equation}
would satisfy this requirement.

\subsection{Fermion Couplings to Electroweak Gauge Bosons}

In order to evaluate precision electroweak observables in our model, we must calculate the strength with which each fermion current couples to
electroweak gauge bosons.  The couplings of boson $V$ to the weak and hypercharge fermion 
(left-handed) currents are given, respectively by the integrals
\begin{equation}
  g^V_W = \int_0^{2\pi R} dy |\psi(y)|^2_W  \chi_V(y)~, \ \ \ \ \ \ \ \  \ \ \ \ g^V_Y = \int_{2\pi R}^{3\pi R} dy |\psi(y)|^2_Y  \chi_V(y)~. 
  \label{eq:couplg}
\end{equation}
Recalling that both the gauge profiles $\chi_{W,Z,\gamma}$ and the
fermion current distributions $|\psi(y)|^2_{W,Y}$ have no
explicit dependence on $g^2_Y$ and  $g_{5Y}^2$ at this order, we anticipate
that the boson-fermion-fermion vertices will be likewise have no explicit dependence
on these couplings.

For the photon, the two integrals are equal and yield
\begin{equation}
  e = 
    \sqrt{2\pi R}\, g_{5W} M_W 
    \left(1-\dfrac{M_W^2}{M_Z^2}\right)^{1/2}
    \left(
      1-\dfrac{2}{3}(M_W \pi R)^2
    \right).
    \label{eq:gggd}
\end{equation}
The $W$ boson couples only  to the fermion $SU(2)$ current (as confirmed by the fact that
$\chi_W(y)$ vanishes for $y > 2\pi R$) and the coupling strength is 
\begin{equation}
  g^W_W = \sqrt{2\pi R}\,  g_{5W} M_W 
    \left(
      1-\dfrac{2}{3}(M_W \pi R)^2
    \right).
    \label{eq:gwwd}
\end{equation}
In a similar manner, $Z$ couplings to the left-handed fermion $SU(2)$ and $U(1)$
currents are 
\begin{equation}
  g^Z_W = \sqrt{2\pi R}\, g_{5W} \dfrac{M_W^2}{M_Z}
    \left(
      1-\dfrac{2}{3}(M_W \pi R)^2
    \right),
    \label{eq:gzwd}
\end{equation}
\begin{equation}
  g^Z_Y = \sqrt{2\pi R}\, g_{5W} \left(\dfrac{M_W^2-M_Z^2}{M_Z}\right)
    \left(
      1-\dfrac{2}{3}(M_W \pi R)^2
    \right),
     \label{eq:gzyd}
\end{equation}
where we note that $g^Z_W > 0$ and $g^Z_Y < 0$
in our phase convention.

For right-handed fermions, the couplings of the photon and $Z$ are
given by the integrals
\begin{equation}
\tilde{g}^V_Y = \int_{2\pi R}^{3\pi R} dy |\widetilde{\psi}(y)|^2_Y  \chi_V(y)~.
\end{equation}
The normalization of this wavefunction, eqn. (\ref{eq:rhnorm}), implies that coupling of 
the photon to the right-handed fermions will be given by $e^2$ of eqn. (\ref{eq:gggd}) --
as required by gauge invariance. From the normalization of the wavefunction, and using
the form of the ideally delocalized left-handed fermion current distribution 
(\ref{eq:hypercurrentdist}), we find
\begin{equation}
g^Z_Y - \tilde{g}^Z_Y = \int_{2\pi R}^{3\pi R} dy\,
\left( |\psi(y)|^2_Y -  |\widetilde{\psi}(y)|^2_Y\right)\,  \chi_Z(y) = {\cal O} \left(e \,M^4_W \pi^4 R^4 \right)~,
\end{equation}
so long as the right-handed current distribution is approximately equal to the left-handed
distribution, eqn. (\ref{eq:lhrhcondition}).


\section{Results in Warped Space: Planck brane $U(1)$ gauge kinetic term}

We turn, now, to considering the case of an $SU(2)\times SU(2) \times U(1)$ gauge
theory in warped space; the fifth dimension is here denoted by coordinate $z$.
The continuum 5D action corresponding to Fig. 1, in conformally flat coordinates,  is given by
\begin{eqnarray}
  S &=& \int_{R}^{R'} dz \int d^4 x \left\{
    -\frac{1}{4g_Y^2} \delta(z-R-0^+) B_{\mu\nu} B_{\rho\lambda}
     \eta^{\mu\rho}\eta^{\nu\lambda}
    \right\}
    \nonumber\\
    & & 
       +\int_R^{R'} dz \int d^4 x \dfrac{R}{z g_{5WA}^2} \left\{
    -\frac{1}{4} W^a_{A\mu\nu} W^a_{A\rho\lambda}
     \eta^{\mu\rho}\eta^{\nu\lambda}
    +\frac{1}{2} W^a_{A\mu z} W^a_{A\nu z}
     \eta^{\mu\nu}
    \right\}
    \nonumber\\
    & & 
       +\int_R^{R'} dz \int d^4 x \dfrac{R}{z g_{5WB}^2} \left\{
    -\frac{1}{4} W^a_{B\mu\nu} W^a_{B\rho\lambda}
     \eta^{\mu\rho}\eta^{\nu\lambda}
    +\frac{1}{2} W^a_{B\mu z} W^a_{B\nu z}
     \eta^{\mu\nu}
    \right\}
    \nonumber\\
    & & 
       +\int_R^{R'} dz \int d^4 x \dfrac{R}{z g_{5Y}^2} \left\{
    -\frac{1}{4} B_{\mu\nu} B_{\rho\lambda}
     \eta^{\mu\rho}\eta^{\nu\lambda}
    +\frac{1}{2} B_{\mu z} B_{\nu z}
     \eta^{\mu\nu}
    \right\},
    \label{eq:action_bulk}
\end{eqnarray}
with $W^a_{A\mu}$ and $W^a_{B\mu}$ being
the bulk $SU(2)$ gauge fields in the $A$- and $B$-branches. 
$B_\mu$ denotes the $U(1)$ gauge field which couples with fermions on either branch.
Note that, in these conformally flat coordinates, one may interpret the action above as having
$z$-dependent couplings (with coupling-squared proportional to $z$) 
in a flat background \cite{SekharChivukula:2005xm}.
We assume large hierarchy between $R$ and $R'$,
\begin{equation}
  R = R' \exp\left(-\frac{b}{2}\right), \qquad
  b \gg 1,
\end{equation}
in order to obtain light $W$ and $Z$ bosons. The operator in the first line of
eqn. (\ref{eq:action_bulk}), which is localized at $z=R$, is a Planck brane hypercharge
kinetic energy term.

It is convenient to define dimensionless bulk gauge couplings,
\begin{equation}
  \dfrac{1}{\tilde{g}_{5WA}^2} \equiv \dfrac{R}{g_{5WA}^2}, \qquad
  \dfrac{1}{\tilde{g}_{5WB}^2} \equiv \dfrac{R}{g_{5WB}^2}, \qquad
  \dfrac{1}{\tilde{g}_{5Y}^2} \equiv \dfrac{R}{g_{5Y}^2}.
\end{equation}
The gauge fields satisfy boundary conditions,
\begin{eqnarray}
  & &
  \partial_z W^a_{A\mu} = 0, 
  \nonumber\\
  & & 
  W^{1,2}_{B\mu} = 0, \quad
  W^{3}_{B\mu} = B_{\mu}, \quad
  \dfrac{1}{\tilde{g}_{5WB}^2} \partial_z  W^{3}_{B\mu} 
  + \dfrac{1}{\tilde{g}_{5Y}^2} \partial_z  B_{\mu} 
  = 0,
\end{eqnarray}
at the $z=R$ boundary, and
\begin{eqnarray}
  & & 
  W^a_{A\mu} = W^a_{B\mu}, \qquad
  \dfrac{1}{\tilde{g}_{5WA}^2} \partial_z  W^a_{A\mu}
  + \dfrac{1}{\tilde{g}_{5WB}^2} \partial_z  W^a_{B\mu} = 0, 
  \nonumber\\
  & & 
  \partial_z  B_{\mu} = 0, 
\end{eqnarray}
at the $z=R'$ boundary.
For simplicity, in the following analyses, 
we assume the bulk gauge couplings in $A$ and $B$
branches are identical,
\begin{equation}
  \tilde{g}_{5W}^2 = \tilde{g}_{5WA}^2 = \tilde{g}_{5WB}^2. 
\end{equation}
The extension to $\tilde{g}_{5WA}^2 \ne \tilde{g}_{5WB}^2$ is 
straightforward. 

\subsection{Mode equations and modified BCs}

The 5D fields $W^a_{A\mu}(x,z)$, $W^a_{B\mu}(x,z)$ and $B_\mu(x,z)$
can be decomposed into KK-modes,
\begin{eqnarray}
  & &
  W_{A\mu}^{1,2}(x,z) = \sum_{n} W_\mu^{(n)1,2}(x) \chi_{W(n)}^A(z),
  \\
  & &
  W_{B\mu}^{1,2}(x,z) = \sum_{n} W_\mu^{(n)1,2}(x) \chi_{W(n)}^B(z),
\end{eqnarray}
and
\begin{eqnarray}
  & &  
  W_{A\mu}^3(x,z) = \gamma_\mu(x) \chi^A_\gamma(z)
    +\sum_{n} Z_\mu^{(n)}(x) \chi^A_{Z(n)}(z), 
  \\
  & &  
  W_{B\mu}^3(x,z) = \gamma_\mu(x) \chi^B_\gamma(z)
    +\sum_{n} Z_\mu^{(n)}(x) \chi^B_{Z(n)}(z), 
  \\
  & &
  B_\mu(x,z) = \gamma_\mu(x) \chi^Y_\gamma(z)
    +\sum_{n} Z_\mu^{(n)}(x) \chi^Y_{Z(n)}(z).
\end{eqnarray}
The mode equations for $W$, $Z$ and $\gamma$ can be read from the 5D
Lagrangian.
They are
\begin{eqnarray}
  & &  
  0 = z \partial_z \left(\frac{1}{z} \partial_z \chi^{A,B}_W(z) \right) 
       + M_W^2 \chi^{A,B}_W(y), 
  \label{eq:wmode_eq_W}
  \\
  & &
  0 = z \partial_z \left(\frac{1}{z} \partial_z \chi^{A,B,Y}_Z(z) \right)
       + M_Z^2 \chi^{A,B,Y}_Z(z), 
  \label{eq:wmode_eq_Z}
  \\
  & &
  0 = z \partial_z \left(\frac{1}{z} \partial_z 
    \chi^{A,B,Y}_\gamma(z) \right).
  \label{eq:wmode_eq_gamma}
\end{eqnarray}
These equations hold at $R+0^+ < z < R'$.
The presence of Planck brane kinetic terms
can be absorbed by the modification of the boundary
conditions, and we find 
\begin{equation}
  0 = \left. 
    \partial_z \chi^A_W(z)\right|_{y=R}, 
\label{eq:bcW1}
\end{equation}
\begin{equation}
  0 = \left. \chi_W^A\right|_{z=R'} - \left. \chi_W^B\right|_{z=R'}, \qquad
  0 = \left. \partial_z \chi_W^A(z) \right|_{z=R'}
     +\left. \partial_z \chi_W^B(z) \right|_{z=R'},
\label{eq:wbcW2}
\end{equation}
\begin{equation}
  0 = \left.\chi^B_W\right|_{z=R}, 
\label{eq:bcW3}
\end{equation}
for the $W$ mode functions, 
\begin{equation}
  0 = \left. 
    \partial_z \chi^A_Z(z) \right|_{z=R},
\label{eq:bcZ1}
\end{equation}
\begin{equation}
  0 = \left. \chi_Z^A\right|_{z=R'} - \left. \chi_Z^B\right|_{z=R'}, \qquad
  0 = \left. \partial_z \chi_Z^A(z) \right|_{z=R'}
     +\left. \partial_z \chi_Z^B(z) \right|_{z=R'},
\label{eq:wbcZ2}
\end{equation}
\begin{equation}
  0 = \left. \chi^B_Z\right|_{z=R} - \left. \chi^Y_Z\right|_{z=R},  
\label{eq:bcZ3a}
\end{equation}
\begin{equation}
  0 = \left. \dfrac{1}{\tilde{g}_{5W}^2} \partial_z \chi^B_Z(z)
      \right|_{z=R}
     +\left. \dfrac{1}{\tilde{g}_{5Y}^2} \partial_z \chi^Y_Z(z)
      \right|_{z=R}
     +\left. M_Z^2 \dfrac{R}{g_Y^2} \chi^Y_Z \right|_{z=R}, 
\label{eq:bcZ3b}
\end{equation}
\begin{equation}
  0 = \left.
    \partial_z \chi^Y_Z(z) 
  \right|_{z=R'}, 
\label{eq:wbcZ4}
\end{equation}
for the $Z$ mode functions, and
\begin{equation}
  0 = \left. 
    \partial_z \chi^A_\gamma(z) \right|_{z=R},
\end{equation}
\begin{equation}
  0 = \left. \chi_\gamma^A\right|_{z=R'} - \left. \chi_\gamma^B\right|_{z=R'}, \qquad
  0 = \left. \partial_z \chi_\gamma^A(z) \right|_{z=R'}
     +\left. \partial_z \chi_\gamma^B(z) \right|_{z=R'},
\end{equation}
\begin{equation}
  0 = \left. \chi^B_Z\right|_{z=R} - \left. \chi^Y_\gamma\right|_{z=R}, 
  \qquad
  0 = \left. \dfrac{1}{\tilde{g}_{5W}^2} \partial_z \chi^B_\gamma(z)
      \right|_{z=R}
     +\left. \dfrac{1}{\tilde{g}_{5Y}^2} \partial_z \chi^Y_\gamma(z)
      \right|_{z=R},
\end{equation}
\begin{equation}
  0 = \left.
    \partial_z \chi^Y_\gamma(z) 
  \right|_{z=R'}, 
\end{equation}
for the photon.

In order to obtain canonically normalized 4D fields ($W$, $Z$, $\gamma$), 
these mode functions are normalized as
\begin{eqnarray}
  & &
  1 = \int_R^{R'} \!\! dz 
      \dfrac{1}{z\tilde{g}_{5W}^2} 
      \left|\chi^A_W(z)\right|^2
     +\int_R^{R'} \!\! dz 
      \dfrac{1}{z\tilde{g}_{5W}^2} 
      \left|\chi^B_W(z)\right|^2,  
  \label{eq:wnormW}
  \\
  & &
  1 = \int_R^{R'} \!\! dz 
      \dfrac{1}{z\tilde{g}_{5W}^2} 
      \left|\chi^A_Z(z)\right|^2
     +\int_R^{R'} \!\! dz 
      \dfrac{1}{z\tilde{g}_{5W}^2} 
      \left|\chi^B_Z(z)\right|^2  
     +\int_R^{R'} \!\! dz 
      \dfrac{1}{z\tilde{g}_{5Y}^2} 
      \left|\chi^Y_Z(z)\right|^2
     +\dfrac{1}{g_Y^2} \left|\chi^Y_Z(R)\right|^2,  
  \nonumber\\
  & &
  \label{eq:wnormZ}
  \\
  & &
  1 = \int_R^{R'} \!\! dz 
      \dfrac{1}{z\tilde{g}_{5W}^2} 
      \left|\chi^A_\gamma(z)\right|^2
     +\int_R^{R'} \!\! dz 
      \dfrac{1}{z\tilde{g}_{5W}^2} 
      \left|\chi^B_\gamma(z)\right|^2  
     +\int_R^{R'} \!\! dz 
      \dfrac{1}{z\tilde{g}_{5Y}^2} 
      \left|\chi^Y_\gamma(z)\right|^2
     +\dfrac{1}{g_Y^2} \left|\chi^Y_\gamma(R)\right|^2 .  
  \label{eq:wnormgamma}
  \nonumber\\
  & &
\end{eqnarray}

\subsection{Gauge Boson and Fermion Profiles}

\subsubsection{$W$ profile}

\label{sec:wprofile}

The mode equation for charged currents, eqn. (\ref{eq:wmode_eq_W}), is
solved by the functions 
\begin{eqnarray}
  f_{WN}(z) 
  &\equiv& 1 - \frac{1}{2}
  (M_W z)^2 \left(\ln\frac{z}{R}-\frac{1}{2}\right)
  +\frac{1}{16} (M_W z)^4 \ln\frac{z}{R} 
  +\cdots,
  \\
  f_{WD}(z)
  &\equiv& M_W z J_1(M_W z)
  \nonumber\\
  &=& \frac{1}{2} (M_W z)^2 -\frac{1}{16} (M_W z)^4 + \cdots. 
\end{eqnarray}
The function $f_{WN}$ satisfies the Neumann condition, eqn. (\ref{eq:bcW3}), at $z=R$, while
$f_{WD}$ satisfies the Dirichlet condition, eqn. (\ref{eq:bcW1}), there.  Hence the $W$ mode
functions can be written 
\begin{equation}
  \chi_W^A(z) = C_W^A f_{WN}(z), 
  \qquad
  \chi_W^B(z) = C_W^B f_{WD}(z), 
\end{equation}
where $C_W^A$ and $C_W^B$ are normalization constants.

As described in appendix \ref{app:warped}, solving for $C^A_W$ and $C^B_W$ 
yields 
\begin{equation}
  C_W^A = \sqrt{\dfrac{2}{b}} \tilde{g}_{5W} \left[
    1 + \frac{3}{16} \frac{2}{b} + \cdots
  \right] = {\frac{2}{b}}\,C^B_W 
  \label{eq:cwab}
\end{equation}

\subsubsection{$Z$ profile}

\label{sec:zprofile}

The $Z$ profile can be studied in a similar manner.
The functions
\begin{eqnarray}
  f_{ZN}(z) 
  &\equiv& 1 - \frac{1}{2}
  (M_Z z)^2 \left(\ln\frac{z}{R}-\frac{1}{2}\right)
  +\frac{1}{16} (M_Z z)^4 \ln\frac{z}{R} 
  +\cdots,
  \\
  f_{ZD}(z)
  &\equiv& M_Z z J_1(M_Z z)
  \nonumber\\
  &=& \frac{1}{2} (M_Z z)^2 -\frac{1}{16} (M_Z z)^4 + \cdots. 
\end{eqnarray}
solve the $Z$ mode differential equation
eqn. (\ref{eq:wmode_eq_Z}). 
The function $f_{ZN}$ satisfies the Neumann condition at $z=R$, while
$f_{ZD}$ satisfies the Dirichlet condition there.

The Neumann condition eqn. (\ref{eq:bcZ1}) at $z=R$  fixes the form of
$\chi_Z^A$, 
\begin{equation}
  \chi_Z^A(z) = C_Z^A f_{ZN}(z), 
\end{equation}
while we express $\chi_Z^B$ and the mode functions in the $Y$ branch as linear combinations of two independent solutions, 
\begin{equation}
  \chi_Z^B(z) = C_Z^B \left[
    f_{ZD}(z) + r_B f_{ZN}(z)
  \right], 
  \end{equation}
\begin{equation}
  \chi_Z^Y(z) = C_Z^Y \left[
    f_{ZD}(z) + r_Y f_{ZN}(z)
  \right], 
\end{equation}
where the Neumann condition at $z=R'$
eqn. (\ref{eq:wbcZ4}) determines the constant $r_Y$,
\begin{equation}
  r_Y = \dfrac{2}{b}.
\end{equation}
As described in Appendix \ref{app:warped}, we can solve for the constant $r_B$
\begin{equation}
  r_B = -\dfrac{s_W^2}{c_W^2 - s_W^2} \dfrac{2}{b}
         \left[
           1 - \frac{3}{8} \dfrac{1}{c_W^2 - s_W^2} \dfrac{2}{b} 
             + \cdots
         \right].
\end{equation}
where
\begin{equation}
  c_W^2 \equiv \dfrac{M_W^2}{M_Z^2}, \qquad s_W^2 = 1 - c_W^2.
\end{equation}
and also solve for the normalization constants $C^A_Z$, $C^B_Z$, and $C_Z^Y$
\begin{equation}
  C^A_Z = \sqrt{\dfrac{2}{b}} \tilde{g}_{5W} c_W \left[
    1 + \dfrac{3}{16}\dfrac{2-c_W^2}{c_W^2}\dfrac{2}{b} + \cdots \right],
    \label{eq:wCAZ}
\end{equation}
\begin{equation}
  C_Z^Y r_Y =   
  C^B_Z r_B = -\sqrt{\dfrac{2}{b}} \tilde{g}_{5W} 
    \dfrac{s_W^2}{c_W} \left[
    1 - \dfrac{3}{16}\dfrac{2}{b} + \cdots \right].
  \label{eq:wCBZ}
\end{equation}

It is important to note that the normalization constants $C_Z^A$ and 
$C_Z^B$ are insensitive to the $U(1)$ couplings $\tilde{g}_{5Y}$ and $g_Y$
individually -- to this order these couplings serve only to split $M^2_W$ from $M^2_Z$.
The mode functions $\chi_Z^A$ and $\chi_Z^B$ are thus identical with
those of a simpler $SU(2)\times SU(2)$ model without bulk
$U(1)$ gauge fields.  This is the same result we observed in the flat space model.

\subsubsection{Photon profile}

\label{sec:gammaprofile}

The photon mode function possesses a flat profile,
\begin{equation}
  \chi^A_\gamma(z) = \chi^B_\gamma(z) = \chi^Y_\gamma(z) = C_\gamma, 
\end{equation}
with $C_\gamma$ being a normalization constant.
The normalization condition eqn. (\ref{eq:wnormgamma}) then reads
\begin{equation}
  1 = C_\gamma^2 \left(
    \dfrac{2}{\tilde{g}_{5W}^2} \dfrac{b}{2}
  + \dfrac{1}{\tilde{g}_{5Y}^2} \dfrac{b}{2}
  + \dfrac{1}{g_Y^2}
    \right).
    \label{eq:normgamma1}
\end{equation}
The RHS is actually independent of $\tilde{g}_{5Y}^2$; as discussed in Appendix \ref{app:warped}, there
is a cancellation between the second and third terms, as may be
seen by inserting eqn. (\ref{eq:gYeff1}) to obtain
\begin{equation}
  C_\gamma = \sqrt{\dfrac{2}{b}} \tilde{g}_{5W} s_W 
    \left(
      1-\dfrac{3}{16}\frac{2}{b}
     +\cdots
    \right).
    \label{eq:Cgamma}
\end{equation}
In other words, the photon profile is the same as in a simpler $SU(2)\times SU(2)$ model without bulk
$U(1)$ gauge fields.

\subsubsection{Ideally delocalized fermions}

\label{sec:idealfermion}

The ideally delocalized weak $SU(2)$ current distribution of the left-handed fermions is
given by
\begin{eqnarray}
  |\psi^A(z)|^2_W &=& C_\psi \frac{1}{z} f_{WN}(z), \\
  |\psi^B(z)|^2_W &=& C_\psi \frac{1}{z} \frac{b}{2} f_{WD}(z).
\end{eqnarray}
These equations are analogous to (\ref{eq:fermprof}), given eqn. (\ref{eq:cwab})
and interpreting the effects of AdS curvature as yielding -- in conformally flat coordinates --
a gauge-coupling squared proportional to $z$.

Here the normalizion constant $C_\psi$ is fixed by
\begin{equation}
  1 = \int_R^{R'} dz \left[ |\psi^A(z)|^2_W +  |\psi^B(z)|^2_W\right]. 
\end{equation}
We find
\begin{equation}
  C_\psi = \frac{2}{b} \left[
    1 + {\cal O}\left(\frac{1}{b^2}\right)
  \right].
\end{equation}

In order to enable a 5D interpretation of delocalized fermion to be made,
the left-handed hypercharge current distribution $|\psi(z)|^2_Y$ is the same
as that of the weak fermion current:
\begin{equation}
  |\psi(z)|^2_Y = C_\psi \frac{1}{z} \left( 
    f_{WN}(z) + \frac{b}{2} f_{WD}(z)\right).
\end{equation}
As in the case of flat space, we will assume in what follows that the right-handed hypercharge
current distribution $|\widetilde{\psi}(z)|^2_Y$  is approximately equal to the left-handed one
\begin{equation}
|\psi(z)|^2_Y - |\widetilde{\psi}(z)|^2_Y = {\cal O}\left({1\over b}\right)~.
\label{eq:lhrhwarpedcondition}
\end{equation}

\subsection{Fermion Couplings to Electroweak Gauge Bosons}

We are now ready to calculate the couplings of the electroweak gauge bosons to an ideally
delocalized fermion.  The $W$ boson couples with the weak $SU(2)$ fermion current as
\begin{equation}
  g^W_W = \int_R^{R'} dz \left[  |\psi^A(z)|_W^2 \chi_W^A(z)
         + |\psi^B(z)|_W^2 \chi_W^B(z)\right] .
\end{equation}
Similarly, the $Z$ boson and photon couplings to the weak $SU(2)$ and
$U(1)$ left-handed hypercharge currents are given by the integrals 
\begin{equation}
  g^{Z,\gamma}_W = \int_R^{R'} dz \left[ |\psi^A(z)|_W^2 \chi_{Z,\gamma}^A(z)
         + |\psi^B(z)|_W^2 \chi_{Z,\gamma}^B(z) \right],
\end{equation}
\begin{equation}
  g^{Z,\gamma}_Y = \int_R^{R'} dz |\psi(z)|_Y^2 \chi_{Z,\gamma}^Y(z).
\end{equation}

Let us start with the photon coupling. 
It is straightforward to see
\begin{equation}
  g^\gamma_W  = g^\gamma_Y = C_\gamma, 
\end{equation}
in accord with electric charge universality.
\begin{equation}
  e = g^\gamma_W  = g^\gamma_Y.
\end{equation}
We thus obtain
\begin{equation}
  e = \sqrt{\dfrac{2}{b}} \tilde{g}_{5W} s_W \left[
    1 - \frac{3}{16}\frac{2}{b} + \cdots \right].
\end{equation}
By using the profiles of the mode functions determined in the previous
sections, it is also straightforward to calculate other couplings.
We find
\begin{eqnarray}
  g^W_W &=& \sqrt{\dfrac{2}{b}} \tilde{g}_{5W} \left[
    1 - \frac{3}{16}\frac{2}{b} + \cdots \right], \\
  g^Z_W &=& \sqrt{\dfrac{2}{b}} \tilde{g}_{5W} c_W \left[
    1 - \frac{3}{16}\frac{2}{b} + \cdots \right], \\
  g^Z_Y &=& -\sqrt{\dfrac{2}{b}} \tilde{g}_{5W} \dfrac{s_W^2}{c_W} \left[
    1 - \frac{3}{16}\frac{2}{b} + \cdots \right].
\end{eqnarray}
Again, we note that these couplings do not depend on $\tilde{g}_{5Y}$ and $g_Y$
individually.

As in the case of flat space, normalization of the right-handed fermion current distribution
implies that the photon coupling to right-handed fermions will be given by $e$, as
required by gauge invariance.
Additionally, so long as the condition of eqn. (\ref{eq:lhrhwarpedcondition}) is
satisfied, we will have near equality of the left- and right-handed couplings of the $Z$ to
the hypercharge current
\begin{equation}
g^Z_Y - \tilde{g}^Z_Y = {\cal O}\left( {e \over b^2}\right)~.
\end{equation}
%


\section{Precision Electroweak Corrections and Equivalence of Models}

Precision electroweak corrections may be compactly defined with reference to the matrix elements for four-fermion processes.  The most general form of the matrix element for four-fermion neutral weak  current processes 
any Higgsless model may be written \cite{Chivukula:2004af,SekharChivukula:2004mu}
\begin{eqnarray}
-{\cal M}_{NC} = e^2 \frac{{\cal Q}{\cal Q}'}{Q^2} 
& + &
\dfrac{(I_3-s^2 {\cal Q}) (I'_3 - s^2 {\cal Q}')}
	{\left(\frac{s^2c^2}{e^2}-\frac{S}{16\pi}\right)Q^2 +
		\frac{1}{4 \sqrt{2} G_F}\left(1-\alpha T +\frac{\alpha \delta}{4 s^2 c^2}\right)
		} 
\label{eq:NC4} \\ \nonumber & \ \ & \\
&+&
\sqrt{2} G_F \,\frac{\alpha \delta}{s^2 c^2}\, I_3 I'_3 
+ 4 \sqrt{2} G_F  \left( \Delta \rho - \alpha T\right)({\cal Q}-I_3)({\cal Q}'-I_3')~,
\nonumber 
\end{eqnarray}
and the corresponding matrix element for charged currents is 
\begin{eqnarray}
  - {\cal M}_{\rm CC}
  =  \dfrac{(I_{+} I'_{-} + I_{-} I'_{+})/2}
             {\left(\dfrac{s^2}{e^2}-\dfrac{S}{16\pi}\right)Q^2
             +\frac{1}{4 \sqrt{2} G_F}\left(1+\frac{\alpha \delta}{4 s^2 c^2}\right)
            }
        + \sqrt{2} G_F\, \frac{\alpha  \delta}{s^2 c^2} \, \frac{(I_{+} I'_{-} + I_{-} I'_{+})}{2}~.
\label{eq:CC3}
\end{eqnarray}
Here $I^{(\prime)}_a$ and ${\cal Q}^{(\prime)}$ are weak isospin and charge
of the corresponding fermion, $\alpha = e^2/4\pi$, $G_F$ is the usual Fermi constant,
and the weak mixing angle (as defined by the on-shell $Z$ coupling) is denoted by $s^2$ 
($c^2\equiv 1-s^2$). The deviations from the standard model are summarized by
the parameters $\alpha S$, $\alpha T$, $\Delta \rho$, and $\alpha \delta$.

The forms of the couplings of the electroweak gauge bosons to the ideally delocalized fermions in the flat-space model analyzed in section 3 imply that all precision electroweak corrections vanish at the order to which we are working -- just as we found in our work on the deconstructed $SU(2)^N \times U(1)$ linear moose \cite{SekharChivukula:2005xm}.   Because 
eqns. (\ref{eq:gggd}), (\ref{eq:gzwd}), and (\ref{eq:gzyd}) yield the relationship 
\begin{equation}
  -e^2 = g^Z_W g^Z_Y 
\label{eq:vanishS}
\end{equation}
the parameter $\alpha S$ vanishes to this order.  Similarly, because eqns. (\ref{eq:gwwd}), (\ref{eq:gzwd}) and (\ref{eq:gzyd}) imply 
\begin{equation}
  \dfrac{(g^W_W)^2}{M_W^2} = \dfrac{(g^Z_W - g^Z_Y)^2}{M_Z^2},
  \label{eq:vanishT}
\end{equation}
we find that $\alpha T = 0$.  By construction, the fermion and $W$ boson profiles are related as in eq. (\ref{eq:fermprof}).  As a result, if we compute the coupling of the fermion weak current to one of 
the higher charged-current KK modes, $W_{(n\geq 1)}$ (cf. eq. (\ref{eq:couplg})), the result
is
\begin{equation}
  g^{W{(n\geq 1)}}_W = \int_0^{2\pi R} \!\! dy \left\{
      \dfrac{1}{g_{5W}^2} 
      +\dfrac{\delta(y-0^+)}{g_0^2} 
      \right\}
      \chi_{W_{(n\geq 1)}} \chi_W(y), 
\end{equation}
which vanishes because the different KK modes of the W are mutually orthogonal.  Hence, exchange of higher $W$ KK modes makes no contribution to $G_F$, meaning $\alpha\delta = 0$.  Finally, as shown in appendix \ref{app:kk}, we find that the contribution of  higher KK-modes to 
$\Delta\rho - \alpha T$ is negligible; since we have already found that $\alpha T = 0$, we conclude that $\Delta\rho$ also vanishes.  Translating \cite{Chivukula:2004af} to the language of Barbieri et al \cite{Barbieri:2004qk}, we have $\hat{S} = \hat{T} = W = Y = 0$.  

Similarly, our results in the warped-space model analyzed in section 4 show that precision electroweak corrections are at most of order ($1/b^2$).  The couplings derived in section 4.3 ensure that the relationships (\ref{eq:vanishS}) and (\ref{eq:vanishT}) are satisified; these guarantee that $\alpha S$ and $\alpha T$ vanish to order $1/b$.  The fermion and $W$ boson profiles are related such that the coupling of the fermion weak current to one of the higher charged-current KK modes $W_{(n\geq1)}$ is of the form
\begin{equation}
  g^{W{(n\geq 1)}}_W = \int_R^{R\prime} \!\! dz \frac{1}{z} C_\psi 
  \left[ \frac{1}{C^A_W} \chi^A_W(z) \chi^A_{W_{(n\geq 1)}}(z)+ 
  \frac{1}{C^B_W} \chi^B_W(z) \chi^B_{W_{(n\geq 1)}}(z) \right].
  \end{equation}
This vanishes due to the mutual orthogonality of the charged-current KK modes.  As in the flat-space case, then we have $\alpha\delta = 0$.

 The five-dimensional models studied here include both a bulk hypercharge gauge group and a brane
 hypercharge kinetic energy term, and also delocalization of the hypercharge properties of the fermions (correlated to the delocalization of their $SU(2)$ properties).  Nonetheless, we have seen explicitly that the profiles (including normalization) of the neutral gauge bosons and their couplings to the ideally delocalized fermions  have no explicit dependence on either the brane or bulk hypercharge 
 couplings. To this order, these couplings
 serve only to split $M^2_Z$ from $M^2_W$. 
 Naturally, all properties of the charged gauge bosons are also independent of hypercharge.
 We conclude that  studies of the  phenomenology of models with ideal delocalization can be made using a simpler higher-dimensional theory: a five-dimensional $SU(2)_A \times SU(2)_B$ gauge theory with hypercharge entering only through a brane kinetic term and ideal fermion delocalization taking place only with regard to $SU(2)$ properties.  This finding is applied directly in our  study 
 \cite{Chivukula:2005ji} of the multi-gauge boson vertices and chiral lagrangian parameters in Higgsless models with  ideal delocalization. 


\section{A Counter-Example: TeV brane $U(1)$ gauge kinetic term}

\begin{figure}[tbhp]
\centering
\includegraphics[width=0.8\textwidth]{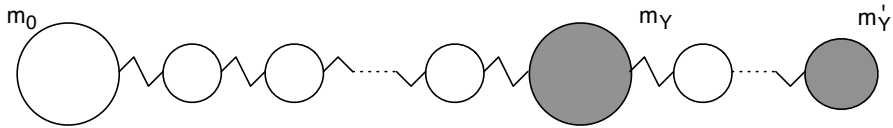}
\label{fig:six}
\caption{Spring system corresponding to the neutral gauge-boson
sector of $SU(2)^2 \times U(1)$ model with a brane kinetic term. $m_0 \propto
1/\tilde{g}^2_{5W}$, $m_Y \propto 1/g^2_Y$, and $m'_Y \propto 1/g'^2_Y$ are larger than all of the other masses in the chain.}
\end{figure}

We now analyze a modified version of our warped-space bulk $SU(2) \times SU(2) \times U(1)$ model which includes  a TeV brane $U(1)$ gauge kinetic term. This model provides  
a counter-example to our previous discussion in the sense that the bulk and brane kinetic $U(1)$
couplings will now appear separately in the calculation of the $\gamma$, $W$, and $Z$ couplings.  
The action for this model\footnote{We should note that the models discussed
in refs. \cite{Cacciapaglia:2004rb,Cacciapaglia:2004jz} envision an $SU(2)_L \times SU(2)_R$
bulk gauge theory, with the left-handed fermion zero-modes arising from bulk fermions charged
under $SU(2)_L$ and the right-handed ones arising from bulk fermions charged under
$SU(2)_R$. In the models discussed in
this paper, the left-handed fermion zero modes arise from bulk fermions  charged under both $SU(2)_A$ and $SU(2)_B$, while the right-handed zero modes arise from bulk fermions charged only
under $U(1)_Y$. Because the $W$ boson arises from both $SU(2)_A$ and $SU(2)_B$, ideal fermion delocalization cannot be realized with a $SU(2)_L \times SU(2)_R$ gauge structure.}
 contains the term \cite{Cacciapaglia:2004jz}
\begin{equation}
  S_{\rm TeV}
  = \int_R^{R'} dz \int d^4 x \left\{
      - \dfrac{1}{4g'^2_Y} \delta(z-R'+0^+) 
        B_{\mu\nu} B_{\rho\lambda} \eta^{\mu\rho} \eta^{\nu\lambda}
    \right\}~,
\label{eq:action_tev}
\end{equation}
in addition to those given in eqn. (\ref{eq:action_bulk}).

The spring system corresponding to the neutral gauge boson sector in this
case is shown in figure 6 and, as drawn, we will consider the
system in the limit that $g^2_Y  <  g'^2_Y$ or $m'_Y  <  m_Y$. The zero mode corresponds
to translation of the entire system, as before. The lightest non-zero mode
again corresponds, roughly, to a ``breathing" mode with masses $m_0$ and $m_Y$
oscillating slowly apart. To the extent $m'_Y$ is not negligible, however, we expect that
the distance between $m_Y$ and $m'_Y$ will oscillate slowly as well. In this case, it is not
possible to replace all of the $U(1)$ masses by a single effective mass -- as in eqn. 
(\ref{eq:tildem}). As we will see, this will lead to potentially large corrections to the electroweak parameters of order $(m'_Y/m_Y)^2$.

The calculations proceed analogously to those in section 4, see
appendix \ref{app:u1brane}. We will perform the calculation
perturbatively in the coupling $g'^2_Y$, so it will be convenient to
define
\begin{equation}
  \eta' \equiv \dfrac{\tilde{g}_{5Y}^2}{g'^2_Y} \dfrac{2}{b}\propto \dfrac{m'_Y}{m_Y}~.
\end{equation}
For ideally delocalized fermions, we find
\begin{eqnarray}
  e     &=& \sqrt{\dfrac{2}{b}} \tilde{g}_{5W} s_W \left[
              1 - \frac{3}{16}\frac{2}{b} 
                + \dfrac{1}{4} \dfrac{s_W^2}{c_W^2} 
                  \dfrac{\tilde{g}_{5W}^2}{\tilde{g}_{5Y}^2} \eta'^2
                + \cdots \right], 
                \label{eq:etagamma}
  \\
  g^W_W &=& \sqrt{\dfrac{2}{b}} \tilde{g}_{5W} \left[
              1 - \frac{3}{16}\frac{2}{b} 
                + \cdots \right], 
  \\
  g^Z_W &=& \sqrt{\dfrac{2}{b}} \tilde{g}_{5W} c_W \left[
              1 - \frac{3}{16}\frac{2}{b} 
                - \dfrac{1}{4} \dfrac{s_W^4}{c_W^4} 
                  \dfrac{\tilde{g}_{5W}^2}{\tilde{g}_{5Y}^2} \eta'^2
                + \cdots \right], 
  \\
  g^Z_Y &=& -\sqrt{\dfrac{2}{b}} \tilde{g}_{5W} \dfrac{s_W^2}{c_W}
             \left[
              1 - \frac{3}{16}\frac{2}{b} 
                - \dfrac{1}{4} \dfrac{s_W^4}{c_W^4} 
                  \dfrac{\tilde{g}_{5W}^2}{\tilde{g}_{5Y}^2} \eta'^2
                + \cdots \right]. 
\end{eqnarray}
Note the non-trivial dependence\footnote{The couplings reproduce
the results of section 4 in the limit $\eta' \to 0$.}
on $\eta'$, and hence dependence on $\tilde{g}_{5Y}$ beyond that
encoded in the splitting between $M^2_W$ and $M^2_Z$.
These couplings result in non-vanishing $\alpha S$ and $\alpha T$,
\begin{eqnarray}
  \alpha S &=& -2 \dfrac{s^4}{c^2} 
                  \dfrac{\tilde{g}_{5W}^2}{\tilde{g}_{5Y}^2}
                  \eta'^2, 
  \\
  \alpha T &=& -\frac{1}{2} \dfrac{s^4}{c^4} 
                  \dfrac{\tilde{g}_{5W}^2}{\tilde{g}_{5Y}^2}
                  \eta'^2,
                  \label{eq:etaT}
\end{eqnarray}
even for the case of ideal delocalization. This is consistent with the results
of \cite{Cacciapaglia:2004jz} for Higgsless models with localized fermions which are not
precisely ``case 1," as defined in \cite{SekharChivukula:2004mu}.

\section{Conclusions}

In this paper we have discussed ideal delocalization of fermions in a bulk 
$SU(2)\times SU(2) \times U(1)$ Higgsless model with a flat or warped extra dimension.
So as to make an extra dimensional interpretation possible, both the weak and
hypercharge properties of the fermions were delocalized, with
the left-handed $U(1)_Y$ current of fermion being correlated with the $SU(2)_W$
current.   We showed that (up to corrections of subleading order)
ideal fermion delocalization yields vanishing precision electroweak
corrections in this continuum model, as found in the corresponding
theory space models based on deconstruction.  Furthermore, we have shown that
the phenomenology of these models is -- to this order -- equivalent to that
of a simpler $SU(2) \times SU(2)$ model. The leading phenomenological 
constraints on Higgsless models with ideal delocalization come from studies of
the constraints arising from deviations of the $ZWW$ vertex, a topic
investigated in \cite{Chivukula:2005ji}.


\acknowledgments

R.S.C. and E.H.S. are supported in part by the US National Science Foundation under
grant  PHY-0354226. M.K. is supported by a MEXT Grant-in-Aid for Scientific Research
No. 14046201.
M.T.'s work is supported in part by the JSPS Grant-in-Aid for Scientific Research No.16540226. H.J.H. is supported by the US Department of Energy grant
DE-FG03-93ER40757.  R.S.C., M.K., E.H.S., and M.T. gratefully acknowledge the hospitality of the Aspen Center for Physics where this work was completed.


\appendix


\section{Explicit Calculations in Warped Space}

\label{app:warped}

This appendix includes the explicit calculations of the normalization constants for the gauge boson wave functions.

\subsection{$W$ profile normalization constants}

It is convenient to define 
\begin{eqnarray}
  f_{WN}(z) 
  &\equiv& 1 - \frac{1}{2}
  (M_W z)^2 \left(\ln\frac{z}{R}-\frac{1}{2}\right)
  +\frac{1}{16} (M_W z)^4 \ln\frac{z}{R} 
  +\cdots,
  \\
  f_{WD}(z)
  &\equiv& M_W z J_1(M_W z)
  \nonumber\\
  &=& \frac{1}{2} (M_W z)^2 -\frac{1}{16} (M_W z)^4 + \cdots. 
\end{eqnarray}
Both functions satisfy differential equation
eqn. (\ref{eq:wmode_eq_W}). 
The function $f_{WN}$ satisfies the Neumann condition at $z=R$, while
$f_{WD}$ satisfies the Dirichlet condition.
By using these functions we can express the $W$ mode functions satisfying the
boundary conditions Eqs.(\ref{eq:bcW1}) and (\ref{eq:bcW3})  as 
\begin{equation}
  \chi_W^A(z) = C_W^A f_{WN}(z), 
  \qquad
  \chi_W^B(z) = C_W^B f_{WD}(z), 
\end{equation}
where $C_W^A$ and $C_W^B$ are normalization constants.
The boundary condition eqn. (\ref{eq:wbcW2}) then reads
\begin{equation}
  0 = \left(
    \begin{array}{cr}
      f_{WN}(z=R')  & -f_{WD}(z=R') \\
      f'_{WN}(z=R') &  f'_{WD}(z=R')
    \end{array}
  \right)
  \left(
    \begin{array}{c}
      C_W^A \\
      C_W^B
    \end{array}
  \right),
\label{eq:bcW2mat}
\end{equation}
with $f'_{WN}$ and $f'_{WD}$ being defined as 
$f'_{WN}(z) \equiv \partial_z f_N(z)$ and 
 $f'_{WD}(z) \equiv \partial_z f_D(z)$. 
In order to obtain non-zero $C_W^{A,B}$ the determinant of the matrix
in eqn. (\ref{eq:bcW2mat}) should vanish,
\begin{equation}
  0 = \dfrac{f'_{WD}(R')}{f'_{WN}(R')} + \dfrac{f_{WD}(R')}{f_{WN}(R')}. 
\label{eq:bcW2det}
\end{equation}
From the definitions of $f_{WN}(z)$ and $f_{WD}(z)$ we can write the
explicit expansions
\begin{eqnarray}
  f_{WN}(R') &=&
      1 - \frac{b}{4}(M_W R')^2 + \frac{1}{4} (M_W R')^2 
        + \frac{b}{32}(M_W R')^4 + \cdots,
  \\
  f_{WD}(R') &=&
      \frac{1}{2}(M_W R')^2 - \frac{1}{16} (M_W R')^4
        + \cdots,
  \\
  R' f'_{WN}(R') &=&
     -\frac{b}{2}(M_W R')^2 + \frac{b}{8} (M_W R')^4  
        +\cdots,
\label{eq:temp1}
  \\
  R' f'_{WD}(R') &=&
     (M_W R')^2 - \frac{1}{4} (M_W R')^4  
        +\cdots~.
\label{eq:temp2}
\end{eqnarray}
It is now straightforward to determine the $W$ mass as a function of
the warp factor $b$ from eqn. (\ref{eq:bcW2det}). 
We obtain 
\begin{equation}
  (M_W R')^2 = \frac{2}{b} \left[
    1 + \frac{3}{8} \frac{2}{b} + \cdots
  \right].
\label{eq:Mwvsb}
\end{equation}
Note here $M_W$ is suppressed by $1/b$.

From Eqs.(\ref{eq:temp1}) and (\ref{eq:temp2}) we see
\begin{equation}
  \dfrac{f'_{WN}(R')}{f'_{WD}(R')}
  = -\frac{b}{2} + {\cal O}\left(\frac{1}{b}\right).
\end{equation}
Comparing this expression with eqn. (\ref{eq:bcW2mat}) we find
\begin{equation}
  C_W^B = C_W^A \left( \frac{b}{2} + {\cal O}\left(\frac{1}{b}\right) \right).
\end{equation}

We now turn to the normalization condition eqn. (\ref{eq:wnormW}).
It is straightforward to show
\begin{equation}
  \int_R^{R'} dz \dfrac{1}{z} \left| f_{WN}(z) \right|^2 
  = \frac{b}{2} - \frac{7}{16} + {\cal O}\left(\frac{1}{b}\right),
\end{equation}
\begin{equation}
  \int_R^{R'} dz \dfrac{1}{z} \left| f_{WD}(z) \right|^2 
  = \frac{1}{16} \left(\dfrac{2}{b}\right)^2 
   +{\cal O}\left(\dfrac{1}{b^3}\right),
\end{equation}
where we used eqn. (\ref{eq:Mwvsb}) so as to express the results solely
in terms of $b$.
Finally, we can calculate the normalization constants $C_W^{A,B}$,
\begin{equation}
  C_W^A = \sqrt{\dfrac{2}{b}} \tilde{g}_{5W} \left[
    1 + \frac{3}{16} \frac{2}{b} + \cdots
  \right] \approx C_W^B \,\frac{2}{b}~.
\end{equation}

\subsection{$Z$ profile normalization constants}

The $Z$ profile can be studied in a similar manner.
We define
\begin{eqnarray}
  f_{ZN}(z) 
  &\equiv& 1 - \frac{1}{2}
  (M_Z z)^2 \left(\ln\frac{z}{R}-\frac{1}{2}\right)
  +\frac{1}{16} (M_Z z)^4 \ln\frac{z}{R} 
  +\cdots,
  \\
  f_{ZD}(z)
  &\equiv& M_Z z J_1(M_Z z)
  \nonumber\\
  &=& \frac{1}{2} (M_Z z)^2 -\frac{1}{16} (M_Z z)^4 + \cdots. 
\end{eqnarray}
Both functions satisfy differential equation
eqn. (\ref{eq:wmode_eq_Z}). 
The function $f_{ZN}$ satisfies the Neumann condition at $z=R$, while
$f_{ZD}$ satisfies the Dirichlet condition.

The Neumann condition, eqn. (\ref{eq:bcZ1}), at $z=R$  fixes the form of
$\chi_Z^A$, 
\begin{equation}
  \chi_Z^A(z) = C_Z^A f_{ZN}(z), 
\end{equation}
while we express $\chi_Z^B$ as a linear combination of two independent
solutions, 
\begin{equation}
  \chi_Z^B(z) = C_Z^B \left[
    f_{ZD}(z) + r_B f_{ZN}(z)
  \right], 
\label{eq:chi_ZB}
\end{equation}
with $r_B$ being a constant.
The mode function in the Y-branch ($\chi_Z^Y$) can also be expressed
as
\begin{equation}
  \chi_Z^Y(z) = C_Z^Y \left[
    f_{ZD}(z) + r_Y f_{ZN}(z)
  \right], 
\label{eq:chi_ZY}
\end{equation}
where the Neumann condition, eqn.  (\ref{eq:wbcZ4}), at $z=R'$
determines the constant $r_Y$,
\begin{equation}
  r_Y = \dfrac{2}{b}.
\label{eq:r_Y}
\end{equation}

The constant $r_B$ is determined from the boundary condition at $z=R'$, eqn. (\ref{eq:wbcZ2}),
 which may be written 
\begin{equation}
  0 = \left(
    \begin{array}{cr}
      f_{ZN}(R')  & -f_{ZD}(R') - r_B f_{ZN}(R')\\
      f'_{ZN}(R') &  f'_{ZD}(R') + r_B f'_{ZN}(R')
    \end{array}
  \right)
  \left(
    \begin{array}{c}
      C_Z^A \\
      C_Z^B
    \end{array}
  \right).
\label{eq:bcZ2mat}
\end{equation}
Again the determinant should vanish. 
We thus find
\begin{equation}
  0 = \dfrac{f'_{ZD}(R')}{f'_{ZN}(R')}
     +\dfrac{f_{ZD}(R')}{f_{ZN}(R')}
     +2r_B. 
\label{eq:bcZ2det}
\end{equation}
It is easy to show
\begin{equation}
  \dfrac{f'_{ZD}(R')}{f'_{ZN}(R')} = -\dfrac{2}{b}.
\label{eq:Zderiv_ratio}
\end{equation}
The calculation of $f_{ZD}(R')/f_{ZN}(R')$ is a little more involved. 
We introduce the weak mixing angle defined by the $M_W/M_Z$ ratio,
\begin{equation}
  c_W^2 \equiv \dfrac{M_W^2}{M_Z^2}, \qquad s_W^2 = 1 - c_W^2.
\end{equation}
Using eqn. (\ref{eq:Mwvsb}), the $M_Z R'$ terms in $f_{ZD}(R')/f_{ZN}(R')$
can be re-expressed in terms of  $b$ and $c_W^2$, yielding 
\begin{equation}
  \dfrac{f_{ZD}(R')}{f_{ZN}(R')}
  = \dfrac{1}{c_W^2 - s_W^2} \dfrac{2}{b} \left[
     1 - \dfrac{3}{4} \dfrac{s_W^2}{c_W^2 - s_W^2} \dfrac{2}{b}  
       + \cdots
    \right].
    \label{eq:rightabove}
\end{equation}
Combining Eqs. (\ref{eq:bcZ2det}), (\ref{eq:Zderiv_ratio}) and (\ref{eq:rightabove}), we find 
\begin{equation}
  r_B = -\dfrac{s_W^2}{c_W^2 - s_W^2} \dfrac{2}{b}
         \left[
           1 - \frac{3}{8} \dfrac{1}{c_W^2 - s_W^2} \dfrac{2}{b} 
             + \cdots
         \right].
\label{eq:r_b}
\end{equation}

This, in turn, leads to a relation between $C^A_Z$ and $C^B_Z$.
From eqn. (\ref{eq:bcZ2mat}) we can read off the equation
\begin{equation}
  0 = C_Z^A + C_Z^B \left( r_B + \dfrac{f'_{ZD}(R')}{f'_{ZN}(R')} \right).
\label{eq:CZ_ab}
\end{equation}
Combining this with  eqn. (\ref{eq:Zderiv_ratio}) and eqn. (\ref{eq:r_b}), we find
\begin{equation}
  C_Z^B r_B = - \dfrac{s_W^2}{c_W^2} \left[
    1 - \dfrac{3}{8 c_W^2} \dfrac{2}{b} + \cdots \right]
  C_Z^A.
\end{equation}

The last piece we need before separately determining $C^A_Z$ and $C^B_Z$ may be obtained by considering the mode function in the $Y$ branch $\chi_Z^Y(z)$,
eqn. (\ref{eq:chi_ZY}).
The boundary condition eqn. (\ref{eq:bcZ3a}) determines the
constant $C_Z^Y$,
\begin{equation}
  C_Z^Y r_Y = C_Z^B r_B = - \dfrac{s_W^2}{c_W^2} \left[
    1 - \dfrac{3}{8 c_W^2} \dfrac{2}{b} + \cdots \right]
  C_Z^A.
\end{equation}
Combining the boundary conditions eqn. (\ref{eq:bcZ3a}) and
eqn. (\ref{eq:bcZ3b}), we also find an expression for the brane kinetic
term $1/g_Y^2$, 
\begin{equation}
  \dfrac{1}{g_Y^2} = 
   -\dfrac{1}{\tilde{g}_{5W}^2} 
    \left. \dfrac{\partial_z \chi_z^B(z)}{M_Z^2 R \chi_Z^B(z)} 
    \right|_{z=R}
   -\dfrac{1}{\tilde{g}_{5Y}^2} 
    \left. \dfrac{\partial_z \chi_z^Y(z)}{M_Z^2 R \chi_Z^Y(z)} 
    \right|_{z=R}.
\label{eq:gY2tmp}
\end{equation}
Plugging eqn. (\ref{eq:chi_ZB}) and eqn. (\ref{eq:chi_ZY}) in
eqn. (\ref{eq:gY2tmp}) we obtain
\begin{equation}
  \dfrac{1}{g_Y^2} = - \dfrac{1}{r_B} \dfrac{1}{\tilde{g}_{5W}^2}
                     - \dfrac{1}{r_Y} \dfrac{1}{\tilde{g}_{5Y}^2}.
\end{equation}
Substituting eqn. (\ref{eq:r_Y}) and eqn. (\ref{eq:r_b}) in this
expression, we further see
\begin{equation}
  \dfrac{1}{g_Y^2} + \dfrac{1}{\tilde{g}_{5Y}^2}\dfrac{b}{2}
  = \dfrac{1}{\tilde{g}_{5W}^2}\dfrac{b}{2}
    \dfrac{c_W^2-s_W^2}{s_W^2}\left[
      1 + \frac{3}{8} \dfrac{1}{c_W^2 - s_W^2} \frac{2}{b} + \cdots
    \right].
\label{eq:gYeff1}
\end{equation}
Note that this particular combination of $g_Y$ and $\tilde{g}_{5Y}$
depends only on the bulk $SU(2)$ coupling.

We are now ready to determine the normalization constant $C_Z^A$ from
eqn. (\ref{eq:wnormZ}).   This equation contains both an explicit dependence
on the bulk $U(1)$ coupling $\tilde{g}_{5Y}$ in its third term and an implicit
dependence on $\tilde{g}_{5Y}$ in its fourth term, through the  $g_Y^2$ value calculated in eqn. (\ref{eq:gYeff1}).   However, these two $\tilde{g}_{5Y}$ dependences cancel 
at the order we are working to, as we shall now see.
First, we note that
\begin{equation}
  \int_R^{R'} dz \frac{1}{z}
  \left|f_{ZN}(z)+\frac{1}{r_Y}f_{ZD}(z)\right|^2 
  = \frac{b}{2} + {\cal O}\left(\frac{1}{b}\right).
\label{eq:int_tmp1}
\end{equation}
The absence of a $b^0$ term in eqn. (\ref{eq:int_tmp1}) reflects the
approximate flatness of the mode function $\chi_Z^Y(z)$.
The third and fourth terms of eqn. (\ref{eq:wnormZ}) therefore yield
\begin{equation}
     \int_R^{R'} \!\! dz 
      \dfrac{1}{z\tilde{g}_{5Y}^2} 
      \left|\chi^Y_Z(z)\right|^2
     +\dfrac{1}{g_Y^2} \left|\chi^Y_Z(R)\right|^2  
  = (C_Z^Y)^2 r_Y^2 \left(
    \dfrac{1}{g_Y^2} 
    + \dfrac{1}{\tilde{g}_{5Y}^2}\dfrac{b}{2}\right). 
    \label{eq:int_tmp2}
\end{equation}
Inserting Eqs. (\ref{eq:r_Y}) and (\ref{eq:gYeff1}) in the RHS of this expression, 
confirms the cancellation of the $\tilde{g}_{5Y}$ dependence at this order.

To complete the calculation, we first compute the remaining terms in  eqn. (\ref{eq:wnormZ}) \begin{equation}
  \int_R^{R'} dz \frac{1}{z} |f_{ZN}(z)|^2
  = \frac{b}{2} - \frac{1}{2}\dfrac{1}{c_W^2} +
    \frac{1}{16}\dfrac{1}{c_W^4}
  + \cdots,
\end{equation}
\begin{equation}
  \int_R^{R'} dz \frac{1}{z}
  \left|f_{ZN}(z)+\frac{1}{r_B}f_{ZD}(z)\right|^2 
  = \frac{b}{2} - \frac{1}{2}\dfrac{1}{s_W^2} +
    \frac{1}{16}\dfrac{1}{s_W^4} 
  +\cdots.
\end{equation}
We then obtain the normalization
constants as
\begin{equation}
  C^A_Z = \sqrt{\dfrac{2}{b}} \tilde{g}_{5W} c_W \left[
    1 + \dfrac{3}{16}\dfrac{2-c_W^2}{c_W^2}\dfrac{2}{b} + \cdots \right],
    \label{eq:CAZ}
\end{equation}
and
\begin{equation}
  C^B_Z r_B = -\sqrt{\dfrac{2}{b}} \tilde{g}_{5W} 
    \dfrac{s_W^2}{c_W} \left[
    1 - \dfrac{3}{16}\dfrac{2}{b} + \cdots \right].
    \label{eq:CBZ}
\end{equation}
We emphasize once again that these normalization constants $C_Z^A$,
$C_Z^B$ are insensitive to the bulk $U(1)$ coupling $\tilde{g}_{5Y}$.
The mode functions $\chi_Z^A$ and $\chi_Z^B$ are thus identical with
those of the simpler $SU(2)\times SU(2)$ model without a bulk
$U(1)$ gauge field.


\section{KK-mode contribution to $\Delta\rho - \alpha T$ in flat space}

\label{app:kk}

In this appendix we restrict ourselves to the $g_Y \to 0$ limit; 
extension to finite $g_Y$ is straightforward.

In the flat-space model we have a series of neutral KK-modes,
\begin{equation}
  \chi_{Z(n)}(y) \propto \left\{
    \begin{array}{lc}
      0  & \mbox{for $y<2\pi R$},\\
      C_{Z(n)}\sin\left( \dfrac{2n-1}{2R} (y-2\pi R) \right)\quad
         & \mbox{for $y>2\pi R$},
    \end{array}
  \right.
\label{eq:kkmode}
\end{equation}
in addition to the neutral KK-modes which are degenerate with the charged KK modes
in the $g_Y\to 0$ limit.
Since the KK-modes of eq.(\ref{eq:kkmode}) overlap with the
hypercharge current distribution eq.(\ref{eq:hypercurrentdist}),
we need to consider possible contributions \cite{SekharChivukula:2004mu} of these KK modes to
$\Delta\rho - \alpha T$, {\it i.e.} to four-fermion processes at low energies.
However, investigating the coupling of these KK-modes to the fermion
$U(1)$ current, we find 
\begin{equation}
  g^{Z(n)}_Y = \dfrac{4}{\pi} 
      \dfrac{1-(-1)^n}{2n-1} C_{Z(n)} (M_W\pi R)^2,
\end{equation}
with
\begin{equation}
  C_{Z(n)} = \sqrt{\dfrac{2 g_{5Y}^2}{\pi R}}.
\end{equation}
Therefore, contributions from
these KK-modes are suppressed by $(M_W \pi R)^4$,
\begin{equation}
  \dfrac{(g^{Z(n)}_Y)^2}{M_{Z(n)}^2}
  \propto (M_W \pi R)^4
\end{equation}
and are negligible to the order we are working.


\section{Calculations with a TeV Brane $U(1)$ gauge kinetic term}

\label{app:u1brane}

In this appendix, we consider the effect of adding a TeV brane $U(1)$ gauge kinetic term
\begin{equation}
  S_{\rm TeV}
  = \int_R^{R'} dz \int d^4 x \left\{
      - \dfrac{1}{4g'^2_Y} \delta(z-R'+0^+) 
        B_{\mu\nu} B_{\rho\lambda} \eta^{\mu\rho} \eta^{\nu\lambda}
    \right\}~.
\label{eq:action_tev_appendix}
\end{equation}
to the action eqn. (\ref{eq:action_bulk}) \cite{Cacciapaglia:2004jz}.
We perform the calculation in an expansion in powers of
\begin{equation}
  \eta' \equiv \dfrac{\tilde{g}_{5Y}^2}{g'^2_Y} \dfrac{2}{b}~.
\end{equation}
In the following calculations, we neglect ${\cal O}(1/b^2)$, 
${\cal O}(\eta'^3)$, ${\cal O}(\eta'/b)$ contributions to the fermion
couplings -- we only retain terms which suffice  to calculate these
couplings, and therefore  $\alpha S$ and $\alpha T$, 
to ${\cal O}(1/b)$ and ${\cal O}(\eta'^2)$.

Because the charged sector of this
model is independent of $\eta'$, both the $W$ profile and the profile of an ideally
delocalized fermion -- and therefore
the results of sections \ref{sec:wprofile} and \ref{sec:idealfermion} --  are unaltered.

The presence of the TeV brane gauge kinetic action
eqn. (\ref{eq:action_tev_appendix}) modifies the boundary condition of $\chi_Z^Y$
eqn. (\ref{eq:wbcZ4}) and the normalization conditions of the  $Z$ and photon
mode functions eqn. (\ref{eq:wnormZ}) and eqn. (\ref{eq:wnormgamma}), 
\begin{equation}
  0 = \left. \dfrac{1}{\tilde{g}_{5Y}^2} \partial_z \chi_Z^Y(z) 
      \right|_{z=R'}
     -M_Z^2 \dfrac{R'}{g'^2_Y} \left. \chi_Z^Y(z)\right|_{z=R'},
  \label{eq:bcZ4a}
\end{equation}
\begin{eqnarray}
  1 &=&
      \int_R^{R'} \!\! dz 
      \dfrac{1}{z\tilde{g}_{5W}^2} 
      \left|\chi^A_Z(z)\right|^2
     +\int_R^{R'} \!\! dz 
      \dfrac{1}{z\tilde{g}_{5W}^2} 
      \left|\chi^B_Z(z)\right|^2  
     +\int_R^{R'} \!\! dz 
      \dfrac{1}{z\tilde{g}_{5Y}^2} 
      \left|\chi^Y_Z(z)\right|^2
  \nonumber\\
  & & \qquad\qquad 
     +\dfrac{1}{g_Y^2} \left|\chi^Y_Z(R)\right|^2  
     +\dfrac{1}{g_Y'^2} \left|\chi^Y_Z(R')\right|^2,  
  \label{eq:normZa}
  \\
  1 &=&
      \int_R^{R'} \!\! dz 
      \dfrac{1}{z\tilde{g}_{5W}^2} 
      \left|\chi^A_\gamma(z)\right|^2
     +\int_R^{R'} \!\! dz 
      \dfrac{1}{z\tilde{g}_{5W}^2} 
      \left|\chi^B_\gamma(z)\right|^2  
     +\int_R^{R'} \!\! dz 
      \dfrac{1}{z\tilde{g}_{5Y}^2} 
      \left|\chi^Y_\gamma(z)\right|^2
  \nonumber \\
  & & \qquad\qquad
     +\dfrac{1}{g_Y^2} \left|\chi^Y_\gamma(R)\right|^2   
     +\dfrac{1}{g_Y'^2} \left|\chi^Y_\gamma(R')\right|^2 .  
  \label{eq:normgammaa}
\end{eqnarray}

The analysis of the $Z$ mode wavefunction proceeds as 
in section \ref{sec:zprofile} and appendix \ref{app:warped}, with some expressions now
depending on $\eta'$. Specifically, 
\begin{equation}
  r_Y = \dfrac{2}{b}\left[
          1 - \eta' + \dfrac{c_W^2-s_W^2}{2c_W^2} \eta'^2 + \cdots
        \right]~,
\end{equation}
eqn. (\ref{eq:gYeff1}) is replaced by
\begin{equation}
  \dfrac{1}{g_Y^2} + \dfrac{1}{\tilde{g}_{5Y}^2}\dfrac{b}{2} 
  +\dfrac{1}{g_Y'^2}
  = \dfrac{1}{\tilde{g}_{5W}^2}\dfrac{b}{2}
    \dfrac{c_W^2-s_W^2}{s_W^2}\left[
      1 + \frac{3}{8} \dfrac{1}{c_W^2 - s_W^2} \frac{2}{b} 
    \right]
   -\dfrac{1}{2c_W^2} \dfrac{1}{\tilde{g}_{5Y}^2}\dfrac{b}{2} \eta'^2 
   +\cdots ~,
\label{eq:gYeff1a}
\end{equation}
eqns.(\ref{eq:int_tmp1}) and (\ref{eq:int_tmp2}) are, respectively,
replaced by 
\begin{equation}
  \int_R^{R'} dz \frac{1}{z}
  \left|f_{ZN}(z)+\frac{1}{r_Y}f_{ZD}(z)\right|^2 
  = \frac{b}{2}\left[
      1  + {\cal O}\left(\dfrac{\eta'}{b}, \dfrac{1}{b^2} \right)
    \right], 
\label{eq:int_tmp1a}
\end{equation}
and
\begin{eqnarray}
\lefteqn{
     \int_R^{R'} \!\! dz 
      \dfrac{1}{z\tilde{g}_{5Y}^2} 
      \left|\chi^Y_Z(z)\right|^2
     +\dfrac{1}{g_Y^2} \left|\chi^Y_Z(R)\right|^2  
     +\dfrac{1}{g_Y'^2} \left|\chi^Y_Z(R')\right|^2  
}\nonumber\\
  & &
  = (C_Z^Y)^2 r_Y^2 \left(
    \dfrac{1}{g_Y^2} 
    + \dfrac{1}{\tilde{g}_{5Y}^2}\dfrac{b}{2}
    + \dfrac{1}{g_Y'^2} 
    + \dfrac{1}{c_W^2} \dfrac{1}{\tilde{g}_{5Y}^2} \dfrac{b}{2} \eta'^2
    + \cdots
    \right). 
\label{eq:int_tmp2a}
\end{eqnarray}
Ultimately, the expression for the normalization constants
eqns. (\ref{eq:wCAZ}) and (\ref{eq:wCBZ}) become, respectively, 
\begin{equation}
  C^A_Z = \sqrt{\dfrac{2}{b}} \tilde{g}_{5W} c_W \left[
    1 + \dfrac{3}{16}\dfrac{2-c_W^2}{c_W^2}\dfrac{2}{b} 
      - \dfrac{1}{4} \dfrac{s_W^4}{c_W^4} 
        \dfrac{\tilde{g}_{5W}^2}{\tilde{g}_{5Y}^2} \eta'^2
      + \cdots \right]~,
\label{eq:CAZa}
\end{equation}
and 
\begin{equation}
  C^B_Z r_B = -\sqrt{\dfrac{2}{b}} \tilde{g}_{5W} 
    \dfrac{s_W^2}{c_W} \left[
    1 - \dfrac{3}{16}\dfrac{2}{b} 
      - \dfrac{1}{4} \dfrac{s_W^4}{c_W^4} 
        \dfrac{\tilde{g}_{5W}^2}{\tilde{g}_{5Y}^2} \eta'^2
      + \cdots \right].
\label{eq:CBZa}
\end{equation}
Note here that the normalization of the $Z$ mode function in $SU(2)$
branches ($C_Z^{A,B}$) is sensitive to the coupling in the $U(1)$
branch ($\tilde{g}_{5Y}$) at the order of $\eta'^2$.

For the $\gamma$ wavefunction, section \ref{sec:gammaprofile},
we find that eqn. (\ref{eq:normgamma1}) and eqn. (\ref{eq:Cgamma}) are
replaced by
\begin{equation}
  1 = C_\gamma^2 \left(
    \dfrac{2}{\tilde{g}_{5W}^2} \dfrac{b}{2}
  + \dfrac{1}{\tilde{g}_{5Y}^2} \dfrac{b}{2}
  + \dfrac{1}{g_Y^2}
  + \dfrac{1}{g_Y'^2}
    \right),
\label{eq:normgamma1a}
\end{equation}
and
\begin{equation}
  C_\gamma = \sqrt{\dfrac{2}{b}} \tilde{g}_{5W} s_W 
    \left(
      1-\dfrac{3}{16}\frac{2}{b}
      + \dfrac{1}{4} \dfrac{s_W^2}{c_W^2} 
        \dfrac{\tilde{g}_{5W}^2}{\tilde{g}_{5Y}^2} \eta'^2
      + \cdots
    \right).
\label{eq:Cgammaa}
\end{equation}
Again observe that the normalization of the $\gamma$ mode function
($C_\gamma$) is sensitive to the coupling of the $U(1)$ branch
($\tilde{g}_{5Y}$) at this order. 

Calculating the couplings of an ideally 
delocalized fermion, we find the results quoted in eqns. (\ref{eq:etagamma}) -- 
(\ref{eq:etaT}).



\end{document}